\input ./style/arxiv-general.cfg
\documentclass[aoas,MSNbibl,nameyear,rotating,seceqn,dvips]{arximspdf}
\makeatletter
   \@ifpackageloaded{graphicx}{}{\usepackage{graphicx}}
\makeatother
\usepackage{dcolumn}

\doi{10.1214/15-AOAS822}
\volume{9}
\issue{2}
\pubyear{2015}
\firstpage{992}
\lastpage{1023}
\docsubty{FLA}
\setattribute{copyright}{owner}{\textup{In the Public Domain}}

\makeatletter
\newcolumntype{d}[1]{D{.}{.}{#1}}
\def\E{\mathbf{ E}}
\def\Var{\operatorname{\mathbf{Var}}}
\makeatother

\begin{document}
\begin{frontmatter}

\title{Examining socioeconomic health disparities using a~rank-dependent R\'{e}nyi index}
\runtitle{Rank-dependent R\'{e}nyi index}

\begin{aug}
\author[A]{\fnms{Makram} \snm{Talih}\corref{}\ead[label=e1]{mtalih@cdc.gov}}
\runauthor{M. Talih}

\affiliation{National Center for Health Statistics}
\address[A]{Office of Analysis and Epidemiology\\
National Center for Health Statistics\\
Centers for Disease Control and Prevention\\
3311 Toledo Road\\
Hyattsville, Maryland 20782\\
USA\\
\printead{e1}}
\end{aug}

%
\received{\smonth{11} \syear{2014}}
%
\revised{\smonth{2} \syear{2015}}

%
\begin{abstract}
The R\'{e}nyi index (RI) is a one-parameter class of indices that
summarize health disparities among population groups by measuring
divergence between the distributions of disease burden and population
shares of these groups. The \emph{rank-dependent} RI introduced in this
paper is a two-parameter class of health disparity indices that also
accounts for the association between socioeconomic rank and health; it
may be derived from a rank-dependent social welfare function. Two
competing classes are discussed and the rank-dependent RI is shown to
be more robust to changes in the distribution of either socioeconomic
rank or health. The standard error and sampling distribution of the
rank-dependent RI are evaluated using linearization and resampling
techniques, and the methodology is illustrated using health survey data
from the U.S. National Health and Nutrition Examination Survey and
registry data from the U.S. Surveillance, Epidemiology and End Results
Program. Such data underlie many population-based objectives within the
U.S. Healthy People 2020 initiative. The rank-dependent RI provides a
unified mathematical framework for eliciting various societal positions
with regards to the policies that are tied to such wide-reaching public
health initiatives. For example, if population groups with lower
socioeconomic position were ascertained to be more likely to utilize
costly public programs, then the parameters of the RI could be selected
to reflect prioritizing those population groups for intervention or treatment.
\end{abstract}
%
%
\end{frontmatter}

\section{Introduction}\label{intro}

The socioeconomic gradient in health outcomes and resulting health
disparities are now well documented in the United States (U.S.) and
elsewhere [\citet
{costa:quevedo:2012,braveman:etal:2010,wilson:2009,who:2008,lynch:etal:2004,krieger:etal:1997}].
Public health programs
can leverage social determinants of health to address health inequities
and improve health outcomes, as discussed in a recent supplement to
\textit{Public Health Reports} [\citet{dean:etal:2013}]. The U.S. Healthy
People 2020 (HP2020) initiative emphasizes the importance of addressing
the social determinants of health and eliminating disparities: two of
its four overarching goals are to ``create social and physical
environments that promote good health for all'' and ``achieve health
equity, eliminate disparities, and improve the health of all groups''
[\citet{HP.gov:2013}].

Improving overall population health while simultaneously striving to
eliminate health disparities is a fundamental public health and social
policy challenge, because interventions designed to improve the health
of individuals may increase disparities between groups and, conversely,
reducing a group's burden of disease may have little impact on overall
population health [\citet
{frohlich:potvin:2008,mechanic:2002,rose:1985}]. Therefore, it is
imperative that measures of health
disparities be explicit about the value judgments and trade-offs that
are inherent to their methodology---for example, choice of reference
for evaluating disparities, relative versus absolute disparities,
attainment (i.e., favorable outcomes) versus shortfall (i.e., adverse
outcomes) inequalities, equally-weighted versus population-weighted
groups, etc. [\citet
{lambert:zheng:2011,harper:etal:2010,erreygers:2009shortfall,keppel:pamuk:lynch:etal:2005,mackenbach:kunst:1997}].

In the context of socioeconomic disparities in health, the slope index
of inequality [\citeauthor{pamuk:1988} (\citeyear{pamuk:1988,pamuk:1985})], the classical
concentration index [\citet{wagstaff:etal:1991}] and the health
achievement index [\citet{wagstaff:2002}] have provided the impetus for
much of the literature on socioeconomic health inequality measures. For
example, the partial concentration index removes the effect of
covariates (e.g., age or sex) that may be correlated with both health
and income but may be irrelevant to policy in that neither their direct
effect on health nor their joint distribution with income can be
altered [\citet{gravelle:2003}]. Further, an intuitive policy-oriented
interpretation of the concentration index ensues from certain
redistribution schemes [\citet{koolman:vanDoorslaer:2004}].

A slope index of inequality consists of the slope of the (weighted)
least-squares regression of health outcomes onto socioeconomic ranking
and is designed to summarize the association between health and
socioeconomic status (SES). Similarly, the classical concentration
index can be written as twice the covariance between socioeconomic rank
and health shares. A health achievement index represents an
equally-distributed level of health equivalent to the population
average but such that all groups achieve the same average outcome.
Those three indices are interrelated; they are reviewed in Section~\ref
{rankG} of this paper.

Even though the concentration index is widely used, due to its simple
formulation and its appeal to policy makers, its shortcomings have come
under intense scrutiny in recent years [\citet{bleichrodt:etal:2012}]
and various options for correcting its behavior, especially when
measuring socioeconomic inequality in a binary health outcome variable,
have been debated [\citet
{kjellsson:gerdtham:2013,wagstaff:2011,erreygers:2009correcting}].

This paper is not intended as a critique of the concentration index.
Instead, it builds on the differential weighting scheme for
socioeconomic groups [\citet{berrebi:silber:1981}] that the
concentration index utilizes and explores a two-parameter alternative
to the concentration index that is derived from R\'{e}nyi divergence
and includes the entropy-based R\'{e}nyi index of \citet{talih:2012} as
a special case. The proposed approach builds a bridge between the
theory of rank-dependent social welfare functions and the information
theoretic evaluation of divergence between probability distributions.
On the one hand, there is an extensive statistical literature on
discrepancy measures, with applications to goodness-of-fit tests,
robust parameter estimation and signal processing; see \citet
{talih:2012} and the references therein. On the other hand, social
welfare theory provides a framework for the measurement and
characterization of socioeconomic inequalities in health [\citet
{erreygers:vanOurti:2011,bleichrodt:vanDoorslaer:2006}], though social
justice principles remain foundational in socioeconomic inequality
measurement [\citet{bommier:stecklov:2002,peter:2001}].

In parallel with the development of rank-dependent inequality indices,
there is renewed interest in composite indices [\citet
{asada:etal:2013}], particularly for analyses and international
comparisons of wellbeing, for example, using the Human Development
Index [\citet{foster:etal:2013,paruolo:etal:2013}]. In the U.S.,
composite measures of health and health-related quality of life remain
core tools for monitoring progress toward the HP2020 goals [\citet
{HP.gov:2013}]. The focus on multidimensional analyses is also
manifested in the development of indices for \textit{multidimensional
inequality} [\citet
{bennett:mitra:2013,decancq:lugo:2009,tsui:1999,maasoumi:1986}].

The R\'{e}nyi index (RI), reviewed in Section~\ref{G}, is a class of
inequality indices, $\{\operatorname{RI}_\alpha:\alpha\geq0\}$,
that is
derived from R\'{e}nyi divergence [\citet{talih:2012}]. The parameter
$\alpha>0$ is an inequality aversion parameter. The RI is invariant to
the choice of the reference used for evaluating disparities. This
invariance property is relevant to HP2020 and related public health
initiatives because, as mentioned previously, the identification of a
reference involves a value judgment and, moreover, can be affected by
statistical reliability [\citet{fr:2011}]. As discussed in Section~\ref
{G}, the well-known generalized entropy (GE) class also can be modified
for reference invariance. Yet, the RI is more robust than its GE-based
counterpart to changes in the distribution of the adverse health outcome.

Section~\ref{rankG} extends the RI to population groups that are
ordered by family income, educational attainment or other SES variables
(or composites thereof) that contribute to the social determinants of
health. A two-parameter rank-dependent RI is proposed in Section~\ref
{rankG0}, $\{\operatorname{RI}^{(\nu)}_\alpha\dvtx  \alpha\geq0, \nu
\geq1\}$,
where increased values of $\alpha> 0$ reflect an increased societal
aversion to (pure) health inequality and increased values of $\nu> 1$
allow groups with lower SES to weigh more heavily than groups with
higher SES. Section~\ref{rankG1} shows how the rank-dependent RI can be
derived from a rank-dependent social welfare function, relating the
proposed index to the Makdissi--Yazbeck two-parameter classes of health
achievement and inequality indices [\citet{makdissi:yazbeck:2012}]; in
turn, those extend the corresponding Wagstaff classes of indices [\citet
{wagstaff:2002}], reviewed in Section~\ref{rankG00}. (In Appendix~\ref
{rankG2}, a ``convenient regression'' relates the rank-dependent RI to
the slope index of inequality.) In Section~\ref{rankG3}, the GE class
of indices is modified for rank dependence (and reference invariance).
Simulation results in Section~\ref{sim-studies} provide empirical
evidence that the rank-dependent RI is more robust than either of its
Makdissi--Yazbeck or GE-based counterparts to changes in the
distributions of SES or health outcomes.

Sections~\ref{case-study1} and \ref{case-study2} illustrate the
proposed methodology using data from the U.S. National Health and
Nutrition Examination Survey (NHANES), CDC, NCHS, as well as data from
the U.S. Surveillance, Epidemiology and End Results (SEER) Program,
NIH, NCI. Such health survey and registry data are common for tracking
population-based HP2020 objectives. The standard error and sampling
distribution of the rank-dependent RI for these data are evaluated
using linearization and resampling techniques. Even though progress has
been made in understanding the asymptotic behavior of health inequality
indices [\citet{cowell:etal:2011,aaberge:2005}] and first-order
linearization can be adapted for evaluating the sampling variability of
such indices [see Appendix~\ref{stderrors}, and \citet
{langel:tille:2013}, \citeauthor{borrell:talih:2011}
(\citeyear{borrell:talih:2011,borrell:talih:2012}),
\citet{biewen:jenkins:2006}, and \citet{kakwani:etal:1997}], resampling
methods remain most useful for evaluating statistical significance,
especially with complex survey data [\citet
{talih:2012,chen:etal:2012,cheng:etal:2008,harper:etal:2008,rao:etal:1992,rao:wu:1988}].

\section{R\'{e}nyi index}\label{G}

For a population that is partitioned into $M$ mutually exclusive groups
of sizes $n_1,n_2,\ldots,n_M$, with $n=\sum_{j=1}^M n_j$ and $n_j >0$
for $j=1,2,\ldots,M$, consider the distribution of a particular adverse
health outcome $y_{ij}$ for individual $i$ in group $j$. Findings of
health disparities between groups rest on the comparison of the
aggregate health outcomes $y_{\cdot j}=\sum_{i=1}^{n_j}y_{ij}$,
$j=1,2,\ldots,M$, either to one another or to the total, $y_{\cdot
\cdot
}=\sum_{j=1}^{M}y_{\cdot j}$. Below, $y_{\cdot\cdot}$ is assumed to be
positive (i.e., the outcome of interest is observed) and the average
adverse health outcomes for the groups and the total population are
denoted $\bar{y}_{\cdot j}=y_{\cdot j}/n_j$ and $\bar{y}_{\cdot\cdot
}=y_{\cdot\cdot}/n$, respectively.

\textit{Definition}. Let relative health disparities $r_j$ be
proportional to the groups' average adverse health outcomes:
$r_j\propto
\bar{y}_{\cdot j}$. For any positive group weights $p_j$, define $\bar
{p}_j=p_j/\sum_kp_k$ and $\bar{r}_j=r_j/\sum_k\bar{p}_kr_k$. The R\'
{e}nyi index, which takes values in $[0,+\infty]$, is given by
%
\begin{equation}
\label{renyi} \operatorname{RI}_\alpha= %
\cases{\displaystyle -
\frac{1}{1-\alpha}\ln \Biggl(\sum_{j=1}^M
\bar{p}_j\bar {r}_j^{1-\alpha} \Biggr),&\quad $\mbox{for }
\alpha\neq1,\alpha\geq 0,$\vspace*{2pt}
\cr
\displaystyle -\sum_{j=1}^M
\bar{p}_j\ln\bar{r}_j,&\quad $\mbox{for } \alpha= 1.$}
\end{equation}
Thus, $\operatorname{RI}_\alpha=0$ if $\bar{y}_{\cdot j} \equiv\sum_k \bar{p}_k
\bar
{y}_{\cdot k}$. The expression in (\ref{renyi}) is that of the R\'
{e}nyi divergence between the two probability mass functions $\bar
{p}_j$ and $\bar{q}_j:=\bar{p}_j\bar{r}_j$ [\citet{talih:2012,renyi:1960}].

\textit{Remarks}. The $p_j$ are positive weights that are assigned to
each group. Groups are equally weighted ($p_j=1/M$), population
weighted ($p_j=n_j/n$) or, otherwise, reflect a preference ordering,
such as the socioeconomic weights of Section~\ref{rankG}. The $r_j$ are
relative health disparities, where the reference is the population
average ($r_j=\bar{y}_{\cdot j}/\bar{y}_{\cdot\cdot}$), the least
adverse health outcome ($r_j=\bar{y}_{\cdot j}/\min_k\bar{y}_{\cdot
k}$) or, otherwise, any fixed reference such as a HP2020 target
($r_j=\bar{y}_{\cdot j}/y_\mathrm{target}$). Due to the scale invariance
of the RI, the $r_j$ need only be proportional to the groups' average
adverse health outcomes $\bar{y}_{\cdot j}$ [\citet{talih:2012}].\looseness=-1

When $p_j=n_j/n$, the standardized R\'{e}nyi index, with values in
$[0,1]$, is the (between-group) Atkinson index [\citet{atkinson:1970}],
obtained from
%
\begin{equation}
\label{atkinson} A_\alpha=1-e^{-\operatorname{RI}_\alpha}.
\end{equation}

The RI increases with $\alpha$. With infinite inequality aversion
$\alpha\rightarrow\infty$, the RI is dominated by the population group
with the least adverse health outcome:
\[
\lim_{\alpha\rightarrow\infty}\operatorname{RI}_\alpha=-\ln \Bigl(\min
_{1\leq
{k}\leq{M}}\bar{r}_k \Bigr)=:\operatorname{RI}_\infty.
\]
Because $0\leq{A_\alpha}\leq{A_\infty}\leq1$, an alternative
standardization to that in (\ref{atkinson}) emerges:
\[
\frac{A_\alpha}{A_\infty}=\frac{1-e^{-\operatorname{RI}_\alpha
}}{1-e^{-\operatorname
{RI}_\infty}}.
\]

Some of the most commonly used (between-group) health inequality
indices belong to the generalized entropy (GE) class, with values in
$[0,+\infty]$,
\[
\operatorname{GE}_\alpha= \sum_{j=1}^Mp_jg_\alpha(r_j),
\]
where $p_j=n_j/n$, $r_j=\bar{y}_{\cdot j}/\bar{y}_{\cdot\cdot}$, and
%
\begin{equation}
\label{galpha} g_\alpha(r)= %
\cases{\displaystyle \frac{1-r^{1-\alpha}}{1-\alpha},& \quad$\alpha
\neq{1},\alpha \geq{0},$\vspace*{2pt}
\cr
-\ln r,&\quad $\alpha=1;$}
\end{equation}
see \citet{talih:2012} and the literature review therein. When $\alpha=
1$, the R\'{e}nyi and GE indices are equal. When $\alpha\neq{1},
\alpha
\geq{0}$, theses indices are related as follows:
%
\begin{equation}
\label{RI-GE} \operatorname{RI}_\alpha= - \frac{1}{1-\alpha}\ln\bigl[1-(1-
\alpha )\operatorname {GE}_\alpha\bigr].
\end{equation}

An important result from \citet{talih:2012} is that $\operatorname
{RI}_\alpha
\leq\operatorname{GE}_\alpha$ for $\alpha> 1$, which entails that, for
$\alpha> 1$, the RI is more robust than the GE index to changes in the
distribution of health outcomes. For example, consider the hypothetical
populations in Table~\ref{case-studies-tab}, which are studied in
Section~\ref{sim-studies} below. With the commonly used parameter value
$\alpha= 2$, the RI increases $35\%$ between populations 1 and 3, from
$0.257$ to $0.348$, whereas the GE increases $42\%$, from $0.293$ to
$0.417$, $1.2$ times the rate of increase of the RI. With $\alpha= 4$,
the RI increases $26\%$ between populations~1 and 3, from $0.567$ to
$0.717$, whereas the GE increases $70\%$, from $1.494$ to $2.534$, over
$2.6$ times the rate of increase of the RI. Figure~\ref{comparisons}
further illustrates the lack of robustness of the rank-dependent GE
compared with the rank-dependent RI for a range of parameter values.
Robustness is especially important for less common adverse health
outcomes because even small absolute differences between groups can
translate into very large relative disparities $r_j$ and, therefore,
large index values. \citet{harper:etal:2010} provide an excellent
outline of the debate regarding absolute versus relative disparities.

\section{Rank dependence and differential weighting}\label{rankG}
The crucial difference between a rank-dependent health disparity index
and a health disparity index that is not rank dependent is that the
former accounts for the association between an exposure (e.g., SES) and
an outcome (e.g., late-stage uterine cervical cancer), whereas the
latter accounts only for inequalities in the outcome variable.

Let the population groups be ranked from lowest to highest SES, with
$n_j>0$. For $j=1,\ldots,M-1$, define rank variables $R_j$ as follows:
%
\begin{equation}
\label{Rj-formula} R_1=\frac{1}{2}\frac{n_1}{n} \quad\mbox{and}\quad
{R_{j+1}}=\frac
{1}{2}\frac{n_{j+1}}{n}+\sum
_{k=1}^j\frac{n_k}{n}.
\end{equation}
By construction, $0<R_j\leq{R_{j+1}}< 1$. For scalar $\nu\geq{1}$,
define
\[
w_\nu(R_j)=\nu(1-R_j)^{\nu-1}.
\]

The rank-dependent R\'{e}nyi index proposed in this paper is derived
from (\ref{renyi}) using the socioeconomic weights $p_j^{(\nu)}=w_\nu
(R_j)p_j$ instead of just $p_j$, as seen in Section~\ref{rankG0} below.
For $\nu>1$, the initial weights $p_j^{(1)}=p_j$ are rescaled according
to the rank of each group: groups with lower SES are weighted more heavily.
In particular,
%
\begin{equation}
\label{differential} w_\nu(R_j) = \biggl(\frac{1-R_j}{R_j}
\biggr)^{\nu-1}\times w_\nu(1-R_j).
\end{equation}
For example, suppose groups are equally weighted to start, that is,
$p_j \equiv1/M$. Then, for $\nu=2$, the socioeconomic weight for a
group at the first quintile of the SES distribution (i.e., with $R_j =
0.20$) would be $4$ times the socioeconomic weight for the
corresponding group at the fourth quintile of the SES distribution.
With $\nu=3$, this factor grows to $4^{\nu-1} = 4^2 = 16$. When groups
are population weighted initially, that is, $p_j = n_j/n$, the effect
of increasing the value of the parameter $\nu$ is not as clear cut.
Still, Figure~\ref{achieve-vs-capacity} shows, for example, that moving
from $\nu=1$ to $\nu=3$ triples the relative weight of the ``poor'' and
more than doubles the relative weight of the ``near poor,'' while
rendering the weight on the ``high income'' group negligibly small
(these groups are defined in Table~\ref{case-studies-tab}). The
selection of the parameter $\nu$, in practice, will vary according to
the context and data. The analyst is advised to explore different
scenarios and, if required, select the parameter $\nu$ that most
closely reflect his/her expectation.

As seen next, the slope index of inequality, the classical
concentration index, and the extended concentration and health
achievement indices all utilize such differential SES weighting as in
(\ref{differential}), either implicitly or explicitly.\looseness=-1

\subsection{Concentration and health achievement indices}\label{rankG00}
Consider the Q--Q plot of the cumulative distribution of health burden
$y_{\cdot j}$ against the SES rank variables $R_j$ defined previously.
The classical concentration index is defined as twice the area between
the resulting Q--Q curve and the diagonal; equivalently, it can be
written as twice the covariance between SES rank and health burden,
which directly relates it to the slope index of inequality; see \citet
{wagstaff:etal:1991}, \citeauthor{pamuk:1988} (\citeyear{pamuk:1985,pamuk:1988}), as well as
Appendix \ref{rankG2}.

With $p_j = n_j/n$, $r_j=\bar{y}_{\cdot j}/\bar{y}_{\cdot\cdot}$,
and a
normalizing constant $W = \sum_j (1-R_j)p_j$, the classical
concentration index, with values in $[-1,+1]$, can be written as
\[
C=1-\frac{2}{W}\sum_{j=1}^M
(1-R_j)p_j r_j.
\]
The index $C$ takes the value $0$ when the aforementioned Q--Q curve
coincides with the diagonal (i.e., when the covariance between SES rank
and health is $0$).

Even though the concentration index $C$ initially appears value
neutral, this latest expression reveals that $C$ is value laden: all
else being equal, the relative disparities $r_j$ for groups with lower
SES (i.e., lower rank $R_j$) are weighted more heavily than those for
groups with higher SES; specifically, $C$ uses $\nu= 2$ in (\ref
{differential}).

To enable the analyst to account more explicitly for such a value
judgment with respect to the differential weighting of the groups,
\citet
{wagstaff:2002} introduced the extended concentration index, defined
for $\nu\geq1$ as
\[
C(\nu)=1-\frac{\nu}{W(\nu)}\sum_{j=1}^M
(1-R_j)^{\nu-1}p_j r_j,
\]
with normalizing constant $W(\nu) = \sum_j (1-R_j)^{\nu-1}p_j$. As
previously, increasing the value of $\nu$ results in increasingly
larger weights placed on the groups with lower SES, whereas groups with
higher SES are assigned increasingly smaller weights. Thus, the
parameter $\nu$ reflects a degree of socioeconomic inequality aversion.

Between-group disparities, as well as the socioeconomic weighting of
the groups, are sensitive to the implicit (or explicit) value judgments
underlying the classical (or extended) concentration index. Moreover,
assessing disparity based solely on an average health burden fails to
account for that burden's association with SES and the extent of
inequality between the lower and higher SES groups. \citet
{wagstaff:2002} introduced a (rank-dependent) health achievement index
to quantify this trade-off between improving population health and
reducing health inequality. The Wagstaff health achievement index is
defined for $\nu\geq1$ as
\[
H(\nu) = \frac{\nu}{W(\nu)}\sum_{j=1}^M
(1-R_j)^{\nu-1}p_j \bar {y}_{\cdot j}.
\]

With $p_j = n_j/n$ and $r_j=\bar{y}_{\cdot j}/\bar{y}_{\cdot\cdot}$,
$H(1) = \bar{y}_{\cdot\cdot}$, the population average, and the
concentration and health achievement indices are related as follows:
%
\begin{equation}
\label{H-C} H(\nu) = \bigl[1 - C(\nu)\bigr]\times\bar{y}_{\cdot\cdot}.
\end{equation}
As before, consider a particular adverse health outcome $y_{ijk}$ for
individual $i$ in SES group $j$ and population $k$, for example,
late-stage uterine cervical cancer by SES within racial/ethnic
population groups in the U.S. If SES was not accounted for [e.g., $\nu
=1$ and $C(1) = 0$], then only the population means would be compared;
for example, the mean $\bar{y}_{\cdot\cdot1}$ for population $1$
might be higher than the mean $\bar{y}_{\cdot\cdot2}$ for population
$2$, signifying a higher cancer burden for population $1$ than for
population $2$ (e.g., $9.0$ versus $6.4$ per $100\mbox{,}000$). On the other
hand, if SES was accounted for [e.g., $\nu> 1$ and $|C(\nu)| > 0$],
and it was ascertained that the two populations had the same value of
$H(\nu)$ (e.g., $8.64$ per 100,000 for both populations), then this
could occur because, say, population 1 had a more equal distribution
across SES groups [Q--Q curve closer to the diagonal, e.g., $C(\nu)
=0.04$], whereas population~2 had a higher burden of disease for the
lower SES groups [Q--Q curve farther from the diagonal, e.g., $C(\nu) = -0.35$].

Similarly for comparisons over time for a single population, the mean
$\bar{y}_{\cdot\cdot}$ could remain unchanged, yet the health
achievement could become worse due to a shift in the SES distribution
of health burden. Incidentally, precisely for this reason, \citet
{chen:etal:2013} caution against causal inference from socioeconomic
health inequality indices such as the slope index of inequality or the
concentration index. Nonetheless, such indices remain useful for
descriptive as well as comparative analyses in large indicator
initiatives, where resource limitations do not always permit in-depth
causal analyses; the HP2020 initiative, for example, houses over 1200
health indicators [\citet{HP.gov:2013}].

\subsection{Rank-dependent R\'{e}nyi index}\label{rankG0}
As stated previously, the rank-depen\-dent RI is derived from (\ref
{renyi}) using the socioeconomic weights $p_j^{(\nu)}=w_\nu(R_j)p_j$.
To better highlight its connection to social evaluation functions in
Section~\ref{rankG1}, we introduce appropriate notation here, and
re-express the rank-dependent RI accordingly.
In addition, to simplify the remainder of this paper, the $p_j$ will,
henceforth, denote the population-weighted group weights $p_j=n_j/n$.
However, identical derivations follow for equally-weighted groups as
well as any other group weights as a starting point $p_j^{(1)}$.

\textit{Notation}.
For $r >0$, let $f_\alpha$ denote the power transform and
$f^{-1}_\alpha
$ its inverse:
%
\begin{equation}\qquad
\label{falpha} f_\alpha(r)= %
\cases{\displaystyle\frac{r^{1-\alpha}}{1-\alpha},
\vspace*{2pt}
\cr
\ln r, } \qquad
{f^{-1}_\alpha(s)}= %
\cases{ \bigl[(1-\alpha)s \bigr]^{{1}/{(1-\alpha)}},&\quad
 $\alpha \neq{1},\alpha\geq{0},$
\vspace*{2pt}
\cr
e^s,&\quad $\alpha=1.$}
\end{equation}
For $\alpha>0$, the function $f_\alpha$ is the generalized logarithm. Define
\begin{eqnarray*}
W_1(\nu)&=&\sum_{j=1}^Mw_\nu(R_j)p_j,\qquad
W_2(\nu)=\sum_{j=1}^Mw_\nu
(R_j)^2p_j,
\\
\bar{w}_\nu(R_j)&=&\frac{w_\nu(R_j)}{W_1(\nu)},
\\
S(\nu ,\alpha )&=&\sum_{j=1}^M
\bar{w}_\nu(R_j)p_jf_\alpha(r_j).
\end{eqnarray*}
Let $\bar{p}_j^{(\nu)}=p_j^{(\nu)}/\sum_kp_k^{(\nu)}$ and $\bar
{r}_j^{(\nu)}=r_j/\sum_k\bar{p}_k^{(\nu)}r_k$. Using this notation,
we have
\[
\label{pj-nu} \bar{p}^{(\nu)}_j=\bar{w}_\nu(R_j)p_j
\quad\mbox{and}\quad \bar{r}^{(\nu
)}_j=\frac{r_j}{S(\nu,0)},
\]
and the rank-dependent R\'{e}nyi index from (\ref{renyi}) is expressed
for all $\alpha\geq{0}$ and $\nu\geq{1}$ as
%
\begin{equation}
\label{rank-renyi} \operatorname{RI}_\alpha^{(\nu)}=-\ln \biggl\{
\frac{f^{-1}_\alpha
[S(\nu,\alpha
)]}{S(\nu,0)} \biggr\}.
\end{equation}

\subsection{Rank-dependent social evaluation function}\label{rankG1}

A two-parameter social evaluation function is given in aggregate form by
%
\begin{equation}
\label{SEF} S^*(\nu,\alpha)=\sum_{j=1}^M
\bar{w}_\nu(R_j)p_jf_\alpha(\bar
{y}_{\cdot j}),
\end{equation}
where $f_\alpha(\bar{y}_{\cdot j}), \alpha>0$, represents society's
evaluation of the group's health burden $\bar{y}_{\cdot j}$ and $\bar
{w}_\nu(R_j)p_j=\bar{p}_j^{(\nu)}$ is the group's socioeconomic weight
[\citet{makdissi:yazbeck:2012}].

\textit{Remark}. The asterisk in $S^*(\nu,\alpha)$ is to distinguish
it from the \textit{relative} measure $S(\nu,\alpha)$ defined previously,
where the social evaluation function $f_\alpha$ was evaluated at the
relative disparities $r_j$ instead of the average health outcomes $\bar
{y}_{\cdot j}$.

In the above, two components of societal evaluation of health are featured:
\begin{longlist}[(ii)]
\item[(i)] A \textit{pure health inequality} component, driven by society's
evaluation of a group's health burden irrespective of its SES
rank---the function $f_\alpha$ in (\ref{falpha}) has constant
relative-inequality aversion $\alpha=-yf''_\alpha(y)/f'_\alpha(y)$
[\citet{cowell:gardiner:1999,pratt:1964}]. In particular, $\nu=1$
results in a social preference function that is indifferent to SES (at
least explicitly, since, directly or indirectly, SES remains a
determinant of health).
\item[(ii)] A \textit{socioeconomic health inequality} component,
driven by
the rank-dependent weighting function $w_\nu(R_j)$---the parameter
$\nu
$ is a socioeconomic health inequality aversion parameter, with
hyperbolic absolute-inequality aversion $(\nu-2)/(1-R)=-w''_\nu
(R)/w'_\nu(R)$ when $\nu>2$. Here, $\alpha=0$ results in a social
preference function that is indifferent to pure health inequalities
(again, at least explicitly), quantifying solely the distribution of
the adverse health outcome along the SES gradient, as in the extended
concentration index of \citet{wagstaff:2002}.
\end{longlist}

A rank-dependent health achievement index is obtained from
$H^*=f_\alpha
^{-1}(S^*)$ in~(\ref{SEF}); it represents an equally-distributed
equivalent level of health such that $S^*$ is equivalent to $f_\alpha
(H^*)$---that is, a hypothetical society in which all groups achieve an
average outcome $\bar{y}_{\cdot j}$ equal to $H^*$. The \citet
{makdissi:yazbeck:2012} health achievement index is expressed as
%
\begin{equation}
\label{achieve} H^*(\nu,\alpha)= %
\cases{\displaystyle\Biggl[\sum
_{j=1}^M\bar{w}_\nu(R_j)p_j
\bar{y}_{\cdot
j}^{1-\alpha} \Biggr]^{{1}/{(1-\alpha)}},&\quad
$\mbox{for } \alpha
\neq {1},\alpha\geq{0},$\vspace*{2pt}
\cr
\displaystyle\exp \Biggl[\sum
_{j=1}^M\bar{w}_\nu(R_j)p_j
\ln\bar {y}_{\cdot j} \Biggr],&\quad $\mbox{for } \alpha=1.$}
\end{equation}
For example, $H^*(1, 0)$ is the population average outcome $\sum_{j=1}^M\bar{p}_j\bar{y}_{\cdot j}=\bar{y}_{\cdot\cdot}$ (when
$p_j=n_j/n$), whereas $H^*(\nu,0)$ is the SES-weighted population
average outcome $\sum_{j=1}^M\bar{p}_j^{(\nu)}\bar{y}_{\cdot j}$. In
addition, the two limiting cases $\alpha\rightarrow\infty$ and $\nu
\rightarrow\infty$ are important for interpretation:
%
\begin{eqnarray}
\label{minmax} H^*(\nu,\infty)&:=&\lim_{\alpha\rightarrow\infty}H^*(\nu,\alpha )=\min
_{1\leq{k}\leq{M}}\bar{y}_{\cdot{k}} \quad\mbox{and}
\nonumber
\\[-8pt]
\\[-8pt]
\nonumber
 {H^*}(\infty, \alpha)&:=&
\lim_{\nu\rightarrow\infty}H^*(\nu,\alpha)=\bar {y}_{\cdot k^*},
\end{eqnarray}
where $k^*= \arg\min_{1\leq{k}\leq{M}}R_k$ is the group with the lowest
SES rank.

As $\nu>1$ increases, more weight is given to the group with the lowest
SES. If the SES gradient in health is positive when groups are ranked
from highest to lowest SES, then the group with the lowest SES will
also have the worst health outcome $\bar{y}_{\cdot{k^*}}\equiv\max_k\bar
{y}_{\cdot{k}}$. Thus, when $\nu\rightarrow\infty$, society's health
achievement becomes only as good as that of its socioeconomically most
disadvantaged [\citet{rawls:1971}]. On the other hand, holding the
parameter $\nu$ constant, health achievement can only be improved at a
progressively steeper cost of nonintervention, as reflected by
increasing $\alpha>0$. In a society that is infinitely averse to
inequality (and that has unlimited resources), all groups achieve the
best group rate $H^*(\nu,\infty)=\min_k\bar{y}_{\cdot{k}}$.

The rank-dependent RI in (\ref{rank-renyi}) provides a unified
mathematical framework for engaging in the aforementioned
considerations. The index $\operatorname{RI}_\alpha^{(\nu)}$ and the
standardized index $A_\alpha^{(\nu)}$ are
%
\begin{equation}
\label{achieve-ratio} \operatorname{RI}_\alpha^{(\nu)}=-\ln \biggl[
\frac{H^*(\nu,\alpha
)}{H^*(\nu
,0)} \biggr] \quad\mbox{and}\quad {A_\alpha^{(\nu)}}=1-
\frac{H^*(\nu
,\alpha)}{H^*(\nu, 0)}.
\end{equation}
In other words, what equations (\ref{achieve-ratio}), as well as
Figure~\ref{achieve-vs-capacity} below, show is that, for each given
value of
$\nu\geq1$, the standardized rank-dependent RI expresses the relative
change that would be required to ``move the needle'' from the {\it
status quo} [e.g., the reference achievement level $H^*(\nu, 0)$, an
SES-weighted population average health burden] to a level of health
achievement that is compatible with societal expectations [achievement
level $H^*(\nu, \alpha)$ for aversion parameter value $\alpha$].

\subsubsection*{Two-parameter extended concentration index}
As in (\ref{H-C}), when\break $p_j=n_j/n$, the two-parameter
extended concentration index [Makdissi and Yazbeck (\citeyear{makdissi:yazbeck:2012})]
%
\begin{equation}
\label{concentration} C(\nu,\alpha)=1-\frac{H^*(\nu,\alpha)}{H^*(1, 0)}=1-\frac{H^*(\nu
,\alpha)}{\bar{y}_{\cdot\cdot}}
\end{equation}
compares the requisite equally-distributed equivalent health level
$H^*(\nu,\alpha)$ to the population average health outcome $\bar
{y}_{\cdot\cdot}$. $C(\nu,\alpha)$ corresponds to the standardized
index $\tilde{A}_\alpha^{(\nu)}$ that would be obtained if one used
$\bar{r}^{(1)}_j=\bar{y}_{\cdot{j}}/\sum_k\bar{p}_k\bar{y}_{\cdot
{k}}\equiv\bar{y}_{\cdot{j}}/\bar{y}_{\cdot\cdot}$ instead of
$\bar
{r}^{(\nu)}_j=\bar{y}_{\cdot{j}}/\sum_k\bar{p}_k^{(\nu)}\bar
{y}_{\cdot
{k}}$ in (\ref{renyi}). However, unlike the standardized index
$A_\alpha
^{(\nu)}$ in (\ref{achieve-ratio}), $C(\nu,\alpha)$ does not remain
nonnegative. $C(\nu,0)$ and $C(2,0)$ are the extended [$C(\nu)$] and
classical ($C$) health concentration indices, respectively; see
Section~\ref{rankG00}. Instead of the population average outcome $\bar
{y}_{\cdot
\cdot}=H^*(1,0)$ as reference for health achievement, the standardized
index $A_\alpha^{(\nu)}$ in (\ref{achieve-ratio}) uses the SES-weighted
average $H^*(\nu,0)$. The relationship between the standardized
rank-dependent RI, the two-parameter extended concentration index and
the extended concentration index is as follows:
\[
1-A_\alpha^{(\nu)} = \frac{1-C(\nu,\alpha)}{1-C(\nu,0)}.
\]

\begin{figure}

\includegraphics{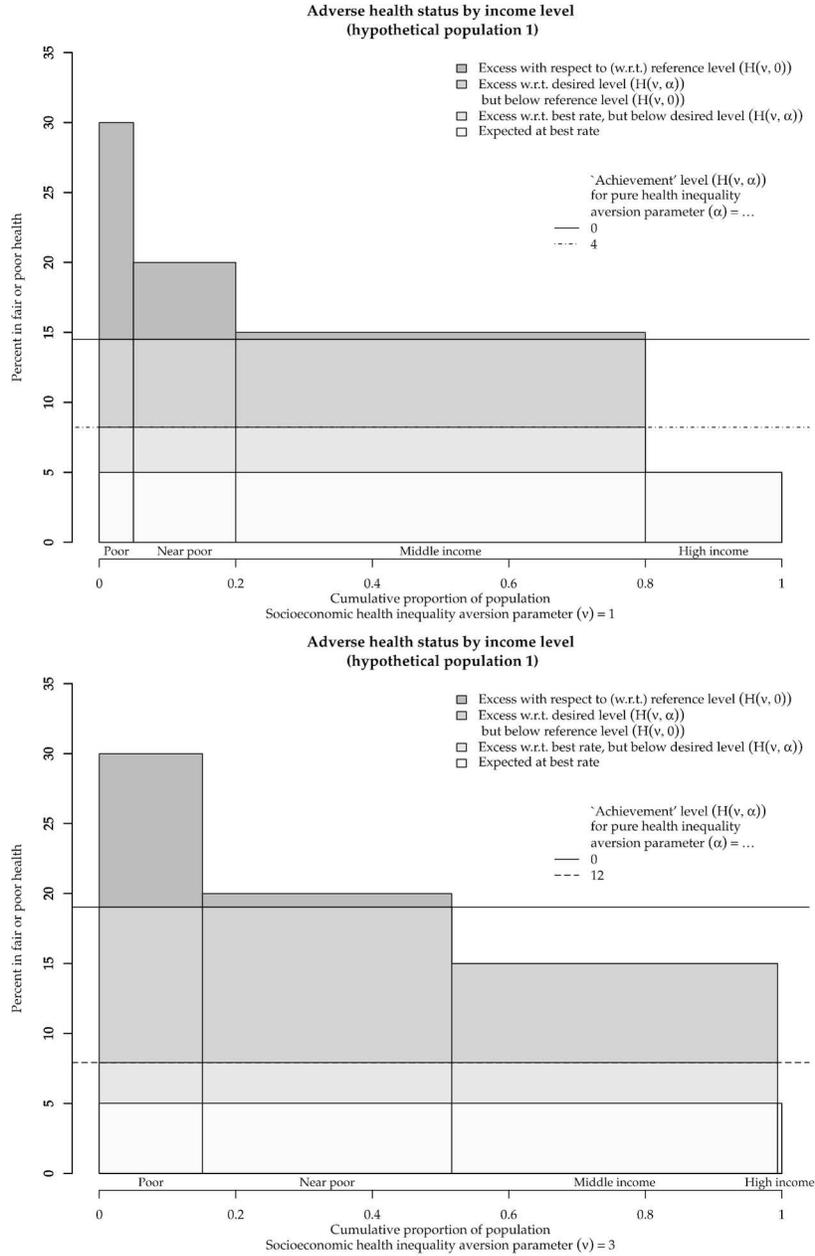}

\caption{Achievement versus capacity to achieve: Illustration using
data for hypothetical population~1 in
Table \protect\ref{case-studies-tab}, with socioeconomic health inequality
parameter $\nu=1$
(rank-neutral group weights; top panel) and $\nu=3$ (weights favorable
to groups with low income level;
bottom panel). The reference ``achievement'' level $H(\nu, 0)$ (solid
lines), that is, the income-weighted
population proportion in fair or poor health, is higher for larger $\nu
$, resulting in a larger gap relative
to the best rate. A larger $\nu$ results in a larger $\alpha$---that
is, a higher
``cost of nonintervention''---for about the same achievement level
$H(\nu, \alpha)\approx8\%$ (dashed lines).} \label{achieve-vs-capacity}
\end{figure}

\subsubsection*{Achievement versus capacity to achieve}
As noted in Section~\ref{G}, the standardization in (\ref{atkinson}) is
not fully satisfactory in that $A_\alpha\rightarrow1$ only if
$\operatorname
{RI}_\alpha\rightarrow\infty$. Thus, for the rank-dependent RI, the
following standardization may be preferable:
%
\begin{equation}
\label{achieve-ratio2} \frac{A_\alpha^{(\nu)}}{A_\infty^{(\nu)}}=\frac{H^*(\nu
,0)-H^*(\nu
,\alpha)}{H^*(\nu,0)-H^*(\nu,\infty)}.
\end{equation}
Holding the parameter $\nu$ constant, $A_\alpha^{(\nu)}/A_\infty
^{(\nu
)}$ is the proportion of the maximum potential improvement in health
achievement [$H^*(\nu,0)-H^*(\nu,\infty)$] that would be attained at
nonintervention cost $\alpha>0$ if all groups were to achieve $\bar
{y}_{\cdot{j}}\equiv{H^*(\nu,\alpha)}$ instead of only $\bar
{y}_{\cdot
{j}}\equiv{H^*(\nu,0)}$.

Figure~\ref{achieve-vs-capacity} illustrates the notion of health
achievement relative to the population's ``capacity to achieve'' using
data for hypothetical population 1 in Table~\ref{case-studies-tab},
with socioeconomic health inequality parameter $\nu=1$ (rank-neutral
group weights) and $\nu=3$ (weights favorable to groups with low income
level). The reference ``achievement'' level $H(\nu,0)$, that is, the
income-weighted population proportion in fair or poor health, is higher
for larger $\nu$, resulting in a larger gap relative to the best rate
$H^*(\nu,\infty)$. A larger $\nu$ results in a larger $\alpha$---that
is, a higher ``cost of nonintervention''---for about the same
achievement level $H(\nu,\alpha)\approx8\%$.
%

\subsection{Rank-dependent generalized entropy class}\label{rankG3}

As stated earlier, the GE index (\ref{galpha}) also can be modified for
rank dependence. Originating in the study of likelihood ratio tests
[\citet{chernoff:1952}], the GE class is tied to important axiomatic
properties in inequality measurement [\citet{cowell:kuga:1981}] and
remains widely used in the economic analysis of income inequalities;
see \citet{talih:2012} for a review of relevant literature.

\begin{table}
\caption{Percentages in fair or poor health by income level for three
hypothetical populations\protect\tabnoteref{TT1}}\label{case-studies-tab}
\begin{tabular*}{\textwidth}{@{\extracolsep{\fill}}lcccc@{}}
\hline
\textbf{Group (}$\bolds{j}$\textbf{}) & \textbf{Poor} & \textbf{Near poor} & \textbf{Middle income} &\textbf{High income}\\
\hline
\textit{Population} 1 & & & &\\
Proportion of population ($n_j/n$) & 0.05 & 0.15 & 0.60 &0.20\\
Percent in fair or poor health ($\bar{y}_{\cdot j})$ & 30\% &20\% &15\%
&5\%\\
\textit{Population} 2 & & & &\\
Proportion of population ($n_j/n$) & 0.05 & 0.15 & 0.60 &0.20\\
Percent in fair or poor health ($\bar{y}_{\cdot j})$ & 30\% &20\% &5\%
&15\%\\
\textit{Population} 3 & & & &\\
Proportion of population ($n_j/n$) & 0.20 & 0.20 & 0.40 &0.20\\
Percent in fair or poor health ($\bar{y}_{\cdot j})$ & 30\% &20\% &15\%
&5\%\\
\hline
\end{tabular*}
\tabnotetext[a]{TT1}{Income level is expressed as a percent of the poverty
threshold. Here, poor${} = {}$below 100\%, near poor${} = {}$100--199\%, middle
income${} = {}$200--399\%, and high income${} = {}$at or above 400\% of the poverty
threshold.}
\end{table}

\textit{Definition}. As before, let the relative health disparities
$r_j$ be proportional to the groups' average adverse health outcomes:
$r_j\propto\bar{y}_{\cdot{j}}$. For any positive group weights $p_j$,
define $\bar{p}_j=p_j/\sum_kp_k$ and $\bar{r}_j=r_j/\sum_k\bar
{p}_kr_k$. A \textit{reference-invariant} GE index is given by
%
\begin{equation}
\label{GE} \operatorname{GE}_\alpha= %
\cases{\displaystyle
\frac{1}{1-\alpha} \Biggl(1-\sum_{j=1}^M
\bar{p}_j\bar {r}_j^{1-\alpha} \Biggr),&\quad $\mbox{for }
\alpha\neq{1},\alpha\geq {0},$\vspace*{2pt}
\cr
\displaystyle -\sum_{j=1}^M
\bar{p}_j\ln\bar{r}_j,&\quad $\mbox{for } \alpha=1.$}
\end{equation}

As before, a \textit{rank-dependent} reference-invariant GE index is
derived from (\ref{GE}) using the socioeconomic weights $p_j^{(\nu
)}=w_\nu(R_j)p_j$ instead of $p_j$. Using the previous notation, the
rank-dependent reference-invariant GE index is expressed for all $\nu
\geq{1}$ and $\alpha\neq{1},\alpha\geq{0}$, as
%
\begin{equation}
\label{rank-GE} \operatorname{GE}_\alpha^{(\nu)}=\frac{1}{1-\alpha}
\biggl\{ 1- \biggl\{\frac
{f^{-1}_\alpha[S(\nu,\alpha)]}{S(\nu,0)} \biggr\}^{1-\alpha} \biggr\}.
\end{equation}
When $\alpha=1$, $\operatorname{GE}_1^{(\nu)}=\operatorname
{RI}_1^{(\nu)}$. For $\alpha
\neq{1}$, the rank-dependent GE index is obtained from the
(standardized) rank-dependent RI as follows, similarly to~(\ref{RI-GE}):
\[
\operatorname{GE}_\alpha^{(\nu)}=\frac{1}{1-\alpha} \bigl\{ 1-
\bigl[1-A_\alpha ^{(\nu)} \bigr]^{1-\alpha} \bigr\}.
\]

As noted earlier, an important result from \citet{talih:2012} is that,
for $\nu\geq{1}$ and $\alpha\geq{1}$, $\operatorname{RI}_\alpha
^{(\nu)}\leq\operatorname
{GE}_\alpha^{(\nu)}$. The inequality is reversed for $0\leq\alpha<1$.
Thus, the rank-dependent RI is more conservative and, therefore, more
robust to changes in the distribution of either SES or health burden
than its GE-based counterpart for \mbox{$\alpha>1$}.

\section{Empirical findings}
\subsection{Simulation studies}\label{sim-studies}
I compare the rank-dependent RI (\ref{rank-renyi}) with the
Makdissi--Yazbeck concentration index (\ref{concentration}) and the
rank-dependent reference-invariant GE index (\ref{rank-GE}) for
hypothetical populations studied by \citet
{keppel:pamuk:lynch:etal:2005}; see Table~\ref{case-studies-tab}.

\begin{sidewaysfigure}

\includegraphics{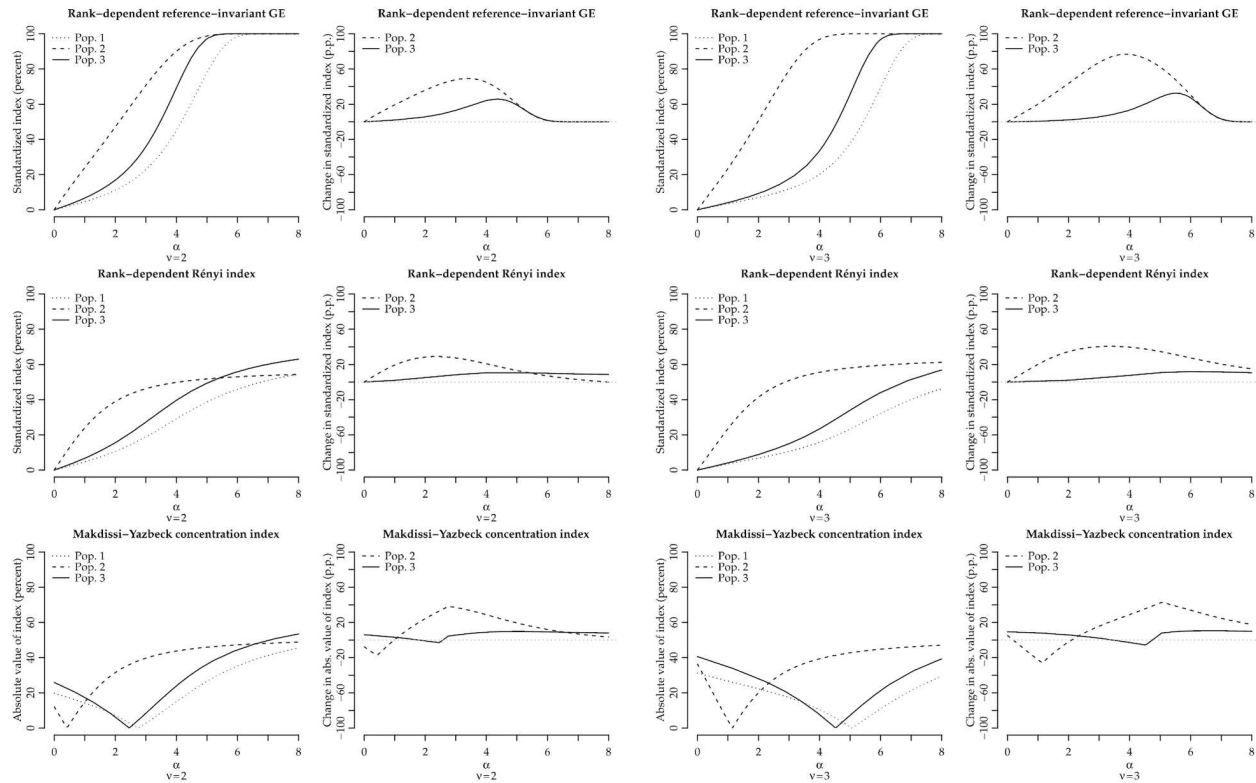}

\caption{Comparison of the rank-dependent R\'{e}nyi index
[(\protect\ref{rank-renyi}) and middle row]
with the Makdissi--Yazbek concentration index [(\protect\ref
{concentration}) and
bottom row] and the
rank-dependent reference-invariant GE index
[(\protect\ref{rank-GE})
and top row]
for hypothetical populations in
Table \protect\ref{case-studies-tab}. For $\nu= 1$, the
rank-dependent R\'
{e}nyi index and the
Makdissi--Yazbek concentration index are equal; therefore, only $\nu=
2$ (two left columns)
and $\nu= 3$ (two right columns) are shown here. By construction, the
rank-dependent R\'{e}nyi
index and rank-dependent reference-invariant GE index are equal to 0
for $\alpha= 0$; the
Makdissi--Yazbek index is not. The class $C(\nu, 0)$ is the Wagstaff
class of extended concentration indices.
For $\nu= 2$, the index $C(2, 0)$ is the ``classical'' health
concentration index.} \label{comparisons}
\end{sidewaysfigure}

In Figure~\ref{comparisons}, the rank-dependent reference-invariant GE
index (top row) is standardized as in (\ref{atkinson}), so that it
takes values in $[0,1]$. In addition, because the Makdissi--Yazbeck index
in (\ref{concentration}) may be negative, only its absolute value is
plotted (bottom row). For $\nu=1$, the rank-dependent RI and the
Makdissi--Yazbeck index are equal; therefore, only $\nu=2$ and $\nu=3$
are shown. By construction, the rank-dependent RI and rank-dependent
reference-invariant GE index are equal to 0 for $\alpha=0$; the
Makdissi--Yazbeck index is not. Conversely, the latter may be zero for
positive values of $\alpha$, whereas the RI and GE index remain
strictly positive unless $\bar{r}_j\equiv{1}$ for all $j$. Shown in the
bottom row of Figure~\ref{comparisons} with $\alpha=0$, the class
$C(\nu
,0)$ is the Wagstaff class $C(\nu)$ of extended concentration indices
and $C(2,0)$ is the classical health concentration index $C$; see
Section~\ref{rankG00}.

The relative ranking of the three hypothetical populations in
Table~\ref{case-studies-tab} changes when the parameters of either of
the three
indices displayed in Figure~\ref{comparisons} are modified. For
example, for the Wagstaff class $C(\nu,0)$, setting $\nu=2$ yields the
ranking $2<1<3$ of the three populations from lowest to highest
inequality, whereas $\nu=3$ results in the ranking $1<2<3$. (Setting
$\nu=4$, not shown, results in the ranking $1<3<2$.) The
Makdissi--Yazbeck index $C(\nu,\alpha)$ further suffers from lack of
smoothness as the pure health inequality aversion parameter $\alpha$
increases, with inequality in some populations assessed to be zero even
for larger values of~$\alpha$. This results in yet further permutations
of the relative ranking of the three hypothetical populations
considered. In contrast, both the rank-dependent RI and rank-dependent
reference-invariant GE index remain smooth functions of the parameter~$\alpha$.
Even though the relative rankings resulting from the use of
either of those two classes of indices are usually in agreement, the
rank-dependent RI is more conservative than its GE-based counterpart
for all values of $\alpha>1$, as known from the inequality at the end
of Section~\ref{rankG3}. As a result, the rankings induced from those
two classes of indices may differ, especially for larger values of
$\alpha$. In addition, the rank-dependent RI is less affected than its
GE-based counterpart by changes to either the health (population 1 vs.
population 2) or income (population~1 vs. population~3) distributions.

\subsection{NHANES case study}\label{case-study1}

During the past 20 years, there was an increase in obesity in the U.S.
Although rates have leveled off in recent years, they remain at
historically high levels. Between 1988--1994 and 2009--2010, the
obesity rate increased 69\% among children and adolescents aged 2--19
years, from 10.0\% to 16.9\% [\citet{ogden:etal:2012}].
%

\begin{table}
\tabcolsep=0pt
\caption{Prevalence of obesity in children and adolescents (aged 2--19
years) by family income, 2001--2010\protect\tabnoteref{TTT1,TTT2}}\label{obesity-tab}
\begin{tabular*}{\textwidth}{@{\extracolsep{\fill}}ld{1.3}d{1.3}d{1.3}d{1.3}d{1.3}@{}}
\hline
\textbf{Income category (}$\bolds{j}$\textbf{)\tabnoteref{TTT3}} &&&&&\\
\textbf{(Family income expressed as}&\textbf{1} &\multicolumn{1}{c}{\textbf{2}} &\multicolumn{1}{c}{\textbf{3}} &\multicolumn{1}{c}{\textbf{4}} &\multicolumn{1}{c}{\textbf{5}}\\
\textbf{percent of poverty threshold)} &
\multicolumn{1}{c}{$\bolds{(<100\%)}$} &
\multicolumn{1}{c}{\textbf{(100--199\%)}}
&
\multicolumn{1}{c}{\textbf{(200--399\%)}}
&
\multicolumn{1}{c}{\textbf{(400--499\%)}}
&
\multicolumn{1}{c}{$\bolds{(\geq500\%)}$}\\
\hline
\textit{NHANES} 2001--2004 & & & & &\\
Prevalence (\%) in group ($\bar{y}_{\cdot j}$) &17.9 &16.7 &17.8 &13.1
&9.8 \\
Standard error (\%)\tabnoteref{TTT4} &1.295 &1.249 &1.182 &1.691
&1.600\\
Population in group ($p_j$)\tabnoteref{TTT5} &0.241 &0.242 &0.291
&0.095 &0.132\\
Rank ($R_j$)\tabnoteref{TTT6} &0.120 &0.362 &0.628 &0.821 &0.934\\[3pt]
\textit{NHANES} 2005--2008 & & & & &\\
Prevalence (\%) in group ($\bar{y}_{\cdot j}$) &19.9 &18.2 &16.0 &14.3
&9.8 \\
Standard error (\%) &1.368 &1.447 &1.403 &2.747 &1.838 \\
Population in group ($p_j$) &0.218 &0.223 &0.298 &0.099 &0.162\\
Rank ($R_j$) &0.109 &0.329 &0.590 &0.788 &0.919\\[3pt]
\textit{NHANES} 2009--2010\tabnoteref{TTT2} & & & & &\\
Prevalence (\%) in group ($\bar{y}_{\cdot j}$) &21.6 &17.4 &15.7 &14.2
&11.5\\
Standard error (\%) &1.306 &1.428 &1.437 &2.686 &2.591 \\
Population in group ($p_j$) &0.232 &0.235 &0.274 &0.088 &0.171\\
Rank ($R_j$) &0.116 &0.349 &0.604 &0.785 &0.914\\
\hline
\end{tabular*}
\tabnotetext[a]{TTT1}{Obesity for children and adolescents aged 2--19 years is
defined as body mass index (BMI) at or above the sex- and age-specific
95th percentile from the 2000 CDC Growth Charts for the U.S. [\citet
{troiano:flegal:1998,growthcharts:2002}].}
\tabnotetext[b]{TTT2}{Data are available biennially and come from the National
Health and Nutrition Examination Survey (NHANES), CDC, NCHS. Preferably
four years of data are pooled for analysis when available [\citet
{NHANESguidelines:2013}], but two-year data are used as a placeholder
to provide the latest data available.}
\tabnotetext[c]{TTT3}{Family income is expressed as a percent of the poverty
threshold; missing values are not included in the analysis.}
\tabnotetext[d]{TTT4}{Standard error evaluated by Taylor linearization [\citet
{SUDAAN:2012,SAS:2010}].}
\tabnotetext[e]{TTT5}{Proportions are rounded for table display and may not add up
to exactly 1.000; unrounded values are used in all calculations.}
\tabnotetext[f]{TTT6}{Rank of group in cumulative distribution of population
computed according to (\ref{Rj-formula}).}
\end{table}

Low income children and adolescents are more likely to be obese than
their higher income counterparts [\citet{ogden:etal:2010}]. In
2009--2010, those with family incomes at or above 500\% of the poverty
threshold had the lowest obesity rate, 11.5\% (Table~\ref
{obesity-tab}). Rates that differed significantly from the lowest rate
at the 0.05 level of significance for children and adolescents with
lower family incomes were as follows: 21.6\% for those under the
poverty threshold, nearly twice the lowest rate; 17.4\% for those with
family incomes at 100--199\% of the poverty threshold, about one and a
half times the lowest rate; and 15.7\% for those with family incomes at
200--399\% of the poverty threshold, almost one and a half times the
lowest rate.

\begin{figure}

\includegraphics[scale=0.98]{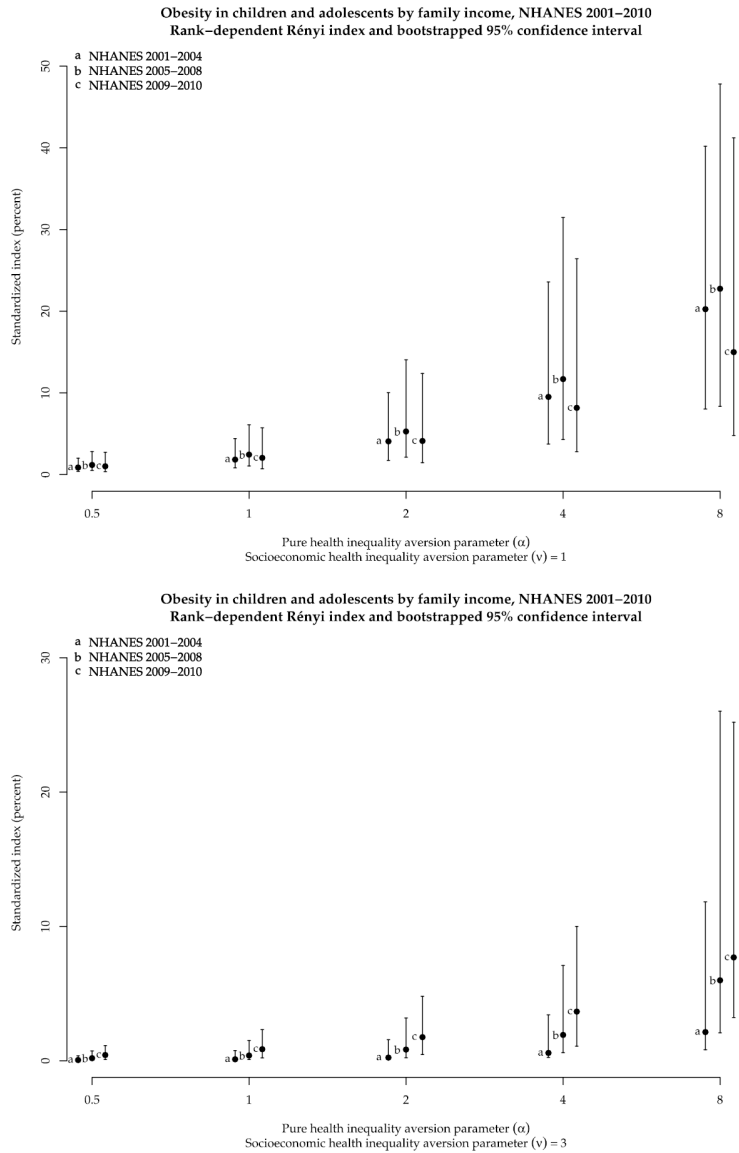}
\vspace*{-3pt}
\caption{Rank-dependent RI and its bootstrapped 95\% confidence
intervals ($B=1000$)
for the prevalence of obesity among children and adolescents aged 2--19
years by family income,
from NHANES 2001--2010 (data in Table \protect\ref{obesity-tab}). For
illustration, values of the
socioeconomic health inequality parameter shown are $\nu=1$
(rank-neutral group weights; top panel)
and $\nu=3$ (weights favorable to those with low family income; bottom
panel).
Values of the pure health inequality aversion parameter shown along the
x-axis are
$\alpha=0.5, 1, 2, 4 $ and $8$. For all combinations of $\nu$ and
$\alpha$ shown, observed differences
between the three survey periods in the rank-dependent RI are not
statistically significant.} \label{obesity-fig}
\end{figure}

The rank variables $R_j$ are computed according to (\ref{Rj-formula})
and shown in Table~\ref{obesity-tab}. Figure~\ref{obesity-fig} displays
the estimated rank-dependent RI together with its bootstrapped 95\%
confidence interval (using $B=1000$ bootstrap samples) for the
prevalence of obesity among children and adolescents by family income,
for NHANES 2009--2010 and the combined cycles 2001--2004 and
2005--2008. For illustration, values of the socioeconomic health
inequality parameter shown in Figure~\ref{obesity-fig} are $\nu=1$
(rank-neutral group weights; top panel) and $\nu=3$ (weights favorable
to those with low family income; bottom panel). Values of the pure
health inequality aversion parameter shown are $\alpha=0.5,1,2,4$ and
$8$. With $\nu=3$, a slight increase in the rank-dependent RI over time
is observed, irrespective of $\alpha$. However, the relative ranking of
the three survey periods changes with $\nu$, as observed in the
simulation studies of Section~\ref{sim-studies} as well as in
Figure~\ref{obesity-fig} for $\nu=1$. Furthermore, for all
combinations of $\nu
$ and $\alpha$ shown, none of the observed differences in the
rank-dependent RI between survey periods are statistically significant
at the 0.05 level of significance.

\textit{Notes}.
Obesity for children and adolescents is defined as body mass index
(BMI) at or above the sex- and age-specific 95th percentile from the
2000 CDC Growth Charts for the U.S. [\citet
{troiano:flegal:1998,growthcharts:2002}]. HP2020 objective NWS-10.4
tracks the proportion of
children and adolescents aged 2--19 years who are considered obese.
Data for NWS-10.4 are from the National Health and Nutrition
Examination Survey (NHANES), CDC, NCHS. Preferably four years of data,
for example, 2009--2012, are pooled [\citet{NHANESguidelines:2013}],
however, at the time of writing this paper, only the two-year data for
2009--2010 were available for analysis.
%
The derivation of a Taylor
linearization approximation of the standard error of the rank-dependent
RI for the various combinations of the parameters $\nu$ and $\alpha$ is
presented in Appendix~\ref{NHANESstderrors}; those standard errors are
used in significance testing for the differences in the rank-dependent
RI between NHANES 2001--2004, 2005--2008 and 2009--2010. The
approximate 95\% confidence intervals shown in Figure~\ref{obesity-fig}
are based on the rescaled bootstrap, which allows the examination of
the sampling distribution of quantities such as the rank-dependent RI
in complex survey data without relying on normality or other
distributional assumptions [\citet
{talih:2012,cheng:etal:2008,rao:etal:1992,rao:wu:1988}].

\subsection{SEER case study}\label{case-study2}

Even though incidence and death rates have declined in recent years for
all cancers, cancer remains a leading cause of death in the U.S.,
second only to heart disease. The cancer objectives for HP2020
underscore the importance of the following: promoting evidence-based
screening for cervical, colorectal and breast cancer in accordance with
U.S. Preventive Services Task Force recommendations; and monitoring the
incidence of invasive (cervical and colorectal) cancer and late-stage
breast cancer, which are intermediate markers of cancer screening
success [\citet{HP.gov:2013}].

For this case study, I examine a subset of the data used to monitor
HP2020 objective C-10, to reduce invasive uterine cervical cancer.
Incidence and treatment of cervical cancer show disparities by race and
ethnicity, SES and health care access [\citet
{saraiya:etal:2013,akers:etal:2007}]. The data in Table~\ref
{cervical-tab} are from the
Surveillance, Epidemiology, and End Results (SEER) Program's 18 Regs
Research Data, NIH, NCI [\citet
{SEERSoft:2013,SEERStat:2013,staging:2001}], and may not be nationally
representative for the U.S.;
see notes below. Nonetheless, for the cases included in Table~\ref
{cervical-tab}, counties where the proportion of persons below the
poverty threshold was lowest (0.00--8.91\%) had the lowest incidence of
invasive uterine cervical cancer, 6.2 cases per 100,000 population (age
adjusted). Rates that differed significantly from the lowest rate at
the 0.05 level of significance for counties with lower area-level SES
were as follows: 8.7 per 100,000 for counties with the highest
proportion (18.87--56.92\%) of persons below the poverty threshold,
nearly one and a half times the lowest rate; 8.0 per 100,000 for
counties with the second highest proportion (14.53--18.86\%) of persons
below the poverty threshold, nearly one and a half times the lowest
rate; and 7.4 per 100,000 for counties with the third highest
proportion (11.61--14.52\%) of persons below the poverty threshold,
19\% higher than the lowest rate.

\begin{table}
\tabcolsep=0pt
\caption{Incidence of invasive uterine cervical cancer (age adjusted,
per 100,000) by area SES, SEER 2006--2010\protect\tabnoteref{TTTT1}}\label{cervical-tab}
{\fontsize{8}{10}\selectfont
\begin{tabular*}{\textwidth}{@{\extracolsep{\fill}}ld{1.3}d{1.3}d{1.3}d{1.3}d{1.3}@{}}
\hline
\textbf{County quintile group (}$\bolds{j}$\textbf{)\tabnoteref{TTTT2}}&&&&& \\
\textbf{(Percentage of persons below}&\multicolumn{1}{c}{\textbf{5}} &\multicolumn{1}{c}{\textbf{4}} &
\multicolumn{1}{c}{\textbf{3}} &\multicolumn{1}{c}{\textbf{2}} &\multicolumn{1}{c}{\textbf{1}}
\\
\textbf{poverty threshold in county)}
&
\multicolumn{1}{c}{\textbf{(18.87--56.92\%)}}
&
\multicolumn{1}{c}{\textbf{(14.53--18.86\%)}}
&
\multicolumn{1}{c}{\textbf{(11.61--14.52\%)}}
&
\multicolumn{1}{c}{\textbf{(8.92--11.60\%)}}
&
\multicolumn{1}{c}{\textbf{(0.00--8.91\%)}}
\\
\hline
{\textit{2006}} & & & & &\\
Incidence in group ($\bar{y}_{\cdot j}$)\tabnoteref{TTTT3} &9.6 &8.9
&7.5 &7.5 &6.4 \\
Standard error\tabnoteref{TTTT4} &0.484 &0.285 &0.339 &0.314 &0.226
\\
Population in group ($p_j$)\tabnoteref{TTTT5} &0.104 &0.267 &0.159
&0.179 &0.292\\
Rank ($R_j$)\tabnoteref{TTTT6} &0.052 &0.237 &0.450 &0.619 &0.854\\
{\textit{2007}} & & & & &\\
Incidence in group ($\bar{y}_{\cdot j}$) &9.0 &9.0 &8.1 &6.9 &6.5 \\
Standard error &0.468 &0.285 &0.353 &0.298 &0.227 \\
Population in group ($p_j$) &0.104 &0.265 &0.160 &0.179 &0.292\\
Rank ($R_j$) &0.052 &0.237 &0.449 &0.618 &0.854\\
{\textit{2008}} & & & & &\\
Incidence in group ($\bar{y}_{\cdot j}$) &9.7 &9.0 &8.0 &7.3 &6.2 \\
Standard error &0.478 &0.285 &0.346 &0.307 &0.220 \\
Population in group ($p_j$) &0.104 &0.263 &0.161 &0.179 &0.293\\
Rank ($R_j$) &0.052 &0.236 &0.448 &0.618 &0.854\\
{\textit{2009}} & & & & &\\
Incidence in group ($\bar{y}_{\cdot j}$) &8.5 &8.4 &7.0 &7.7 &6.4 \\
Standard error &0.450 &0.274 &0.324 &0.316 &0.223 \\
Population in group ($p_j$) &0.104 &0.262 &0.161 &0.179 &0.293\\
Rank ($R_j$) &0.052 &0.236 &0.447 &0.617 &0.853\\
{\textit{2010}} & & & & &\\
Incidence in group ($\bar{y}_{\cdot j}$) &8.7 &8.0 &7.4 &6.4 &6.2 \\
Standard error &0.453 &0.268 &0.330 &0.286 &0.217 \\
Population in group ($p_j$) &0.104 &0.261 &0.162 &0.179 &0.293\\
Rank ($R_j$) &0.052 &0.235 &0.447 &0.617 &0.853\\
\hline
\end{tabular*}}
\tabnotetext[a]{TTTT1}{Data are from the SEER 18 Regs Research Data, NIH, NCI
[\citet{SEERStat:2013,staging:2001}], and are age adjusted using the year
2000 U.S. standard population. Data shown here do not include the full
set of registries used in HP2020 to track objective C-10; thus, data
may not be nationally representative.}
\tabnotetext[b]{TTTT2}{Area socioeconomic status (SES) is computed using
county-level data for the 3141 counties in the year 2000 U.S. Census.
Cutpoints for each SES group, displayed in the table header row,
correspond to the county quintiles when these are sorted according to
the percentage of persons living below the poverty threshold in the county.}
\tabnotetext[c]{TTTT3}{New cases of invasive uterine cervical cancer (age adjusted)
per 100,000 population in group.}
\tabnotetext[d]{TTTT4}{Standard error evaluated by Taylor linearization [\citet
{SEERSoft:2013}].}
\tabnotetext[e]{TTTT5}{Proportions are rounded for table display and may not add up
to exactly 1.000; unrounded values are used in all calculations.}
\tabnotetext[f]{TTTT6}{Rank of group in cumulative distribution of population
computed according to (\ref{Rj-formula}).}
\end{table}

\begin{figure}[p]

\includegraphics{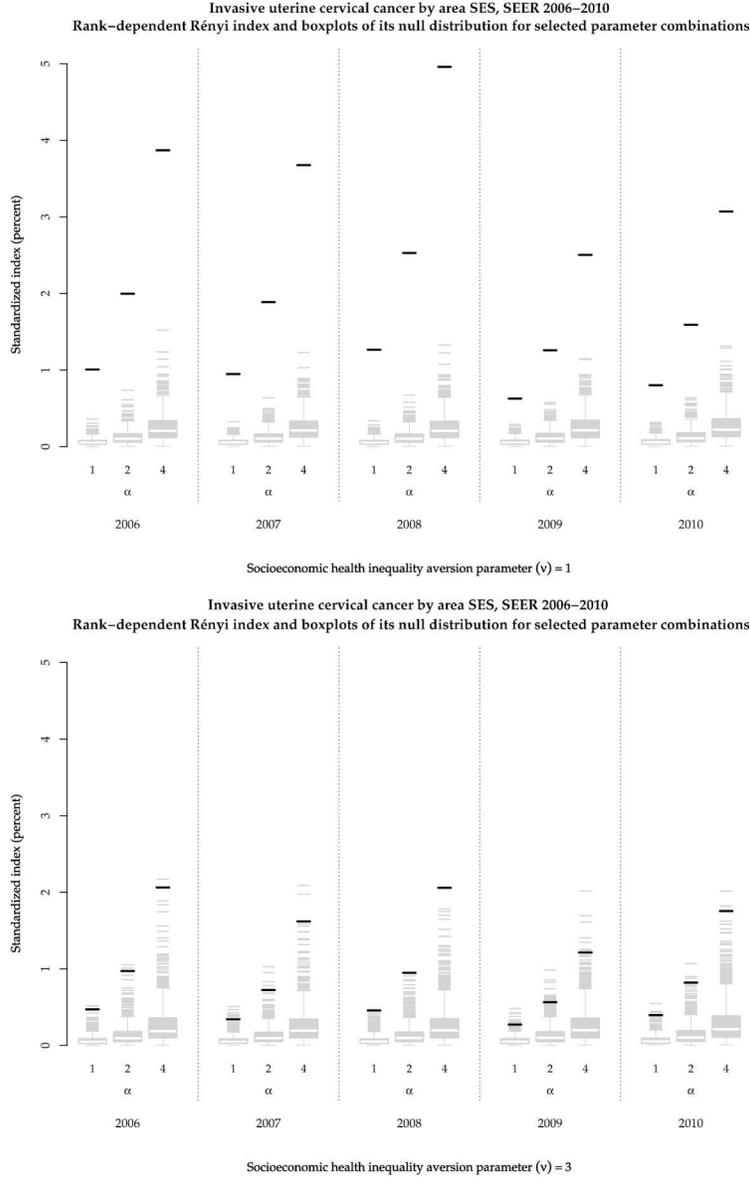}

\caption{Rank-dependent RI and boxplots of its bootstrapped sampling
distribution ($B=1000$) under the null hypothesis of independence for
the incidence of invasive uterine cervical cancer by area SES,
2006--2010, from the SEER Program's 18 Regs Research Data (data in
Table \protect\ref{cervical-tab}). For illustration, values of the
socioeconomic health inequality parameter shown are $\nu=1$
(rank-neutral group weights; top panel) and $\nu=3$ (weights favorable
to groups with low area SES; bottom panel). Values of the pure health
inequality aversion parameter shown are $\alpha=1, 2$ and $4$. For all
combinations of $\nu$ and $\alpha$ shown, the observed rank-dependent
RI differs significantly from its expected value under the null
hypothesis, indicating that the latter can be rejected. However,
changes in the index over time are not statistically significant.}\label{cervical-fig}
\end{figure}

The rank variables $R_j$ are computed according to (\ref{Rj-formula})
and shown in Table~\ref{cervical-tab}. Figure~\ref{cervical-fig}
displays the estimated rank-dependent RI and the boxplot for its
bootstrapped sampling distribution (using $B=1000$ bootstrap samples)
under the null hypothesis of independence for the incidence of invasive
uterine cervical cancer by area SES, 2006--2010, from the SEER 18 Regs
Research Data. For illustration, values of the socioeconomic health
inequality parameter shown in Figure~\ref{cervical-fig} are $\nu=1$
(rank-neutral group weights; top panel) and $\nu=3$ (weights favorable
to groups with low area SES; bottom panel). Values of the pure health
inequality aversion parameter shown are $\alpha=1,2$ and $4$. For all
combinations of $\nu$ and $\alpha$ shown, the observed rank-dependent
RI differs significantly from its expected value under the null
hypothesis, indicating that the latter can be rejected. However,
%
%
for all combinations of $\nu$ and $\alpha$ shown in Figure~\ref
{cervical-fig}, none of the changes in the index over time are
statistically significant at the 0.05 level of significance.

\textit{Notes.}
U.S. cancer registries do not track individual or family income;
therefore, area-level socioeconomic characteristics, linking cancer
cases to U.S. counties, are used to get a proxy for individual-level
SES [\citet{yin:etal:2010,harper:etal:2008}]. In addition to using data
from the SEER Program, HP2020 objective C-10 also uses data collected
through the National Program of Cancer Registries (NPCR), CDC, NCCDPHP.
However, NPCR data are not as readily linked to county-level attributes
as the SEER research data are; the latter are processed via online
queries submitted securely using the SEER*Stat software [\citet
{SEERSoft:2013}]. Further, because cases are linked to counties from
the year 2000 U.S. Census, the analysis does not take into account
changes in county boundaries and/or composition over time. For these
reasons, data and results presented here may not be nationally
representative for the U.S.; they are intended for illustration
purposes only. The derivation of a Taylor linearization approximation
of the standard error of the rank-dependent RI for the various
combinations of the parameters $\nu$ and $\alpha$ is presented in
Appendix \ref{SEERstderrors}; those standard errors are used in
significance testing for the differences in the rank-dependent RI over
time. Because of the assumption of a Poisson distribution for crude
rates, random draws are readily generated under the null hypothesis of
independence. The resulting bootstrapped null distribution for the
rank-dependent RI for each year and combination of the parameters $\nu$
and $\alpha$ is summarized using a boxplot in Figure~\ref{cervical-fig}.

\section{Discussion}\label{DISC}
The rank-dependent RI introduced in this paper is a two-parameter class
of socioeconomic health inequality indices, $\{\operatorname{RI}^{(\nu
)}_\alpha
\dvtx \alpha\geq{0},\nu\geq{1}\}$, where $\alpha>0$ is a constant
relative-inequality aversion parameter and increasing values of the
socioeconomic health inequality aversion parameter $\nu>1$ allow groups
with lower SES gradient to weigh more heavily than groups with higher
SES. In relation to competing index classes such as the
Makdissi--Yazbeck two-parameter extended concentration index and the
rank-dependent reference-invariant GE class, the rank-dependent RI is
more robust to changes in the distribution of either SES or adverse
health outcomes. The proposed method is applicable to a wide range of
public health measures and data, and statistical inference for the
rank-dependent RI is readily implemented using standard statistical software.

The proposed methods are easily extended into a multivariate setting.
As mentioned earlier in the context of the partial concentration index
[\citet{gravelle:2003}], it may be of interest to adjust for covariates
when looking at disparities in health outcomes to rule out those parts
of the SES disparity that might be considered ``just'' or that,
otherwise, cannot be amenable to policy. As an example, if communities
in lower SES are relatively older and have higher rates of cancer for
that reason, findings of socioeconomic disparities might be attenuated
by age, so adjusting for age is of importance. Neighborhood-level or
regional variation may be important for certain outcomes. For example,
illnesses such as influenza outbreaks should adjust for region when
measuring disparities if the outbreak is worse in certain areas of the
country. The SEER data in Section~\ref{case-study2}, above, are age
adjusted. One could also apply the proposed methodology to adjusted
rates obtained from log-linear or logistic regression models. For
example, \citet{rossen:talih:2014} apply the (symmetrized) RI to
population groups obtained from propensity score subclassification,
accounting for demographic and contextual variables to examine
disparities in weight among U.S. children and adolescents.

SES is a multidimensional construct that includes wealth, income,
education and occupation [\citet{talih:2013,krieger:etal:1997}]. Income
and education are used in this paper as univariate SES measures only
for illustration purposes. The proposed methods can also be applied to
the ranking induced from any other SES measure, including composite SES
measures. Nonetheless, the analyst should keep in mind that measuring
occupation as an element of SES remains challenging. Historically,
several approaches have been used, including the Nam--Powers
occupational scale score [\citet{boyd:nam:2004}], which views occupation
as a reflection of education, skill, income and social status, as well
as the National Opinion Research Center's General Social Survey
occupational prestige score [\citet{nakao:treas:1990}], which views
occupation as an indicator of prestige. On the other hand, the O*NET
work content model [\citet{onet:2014}] can provide relevant information
for the measurement of socioeconomic status and whether certain
occupations lead to reduced workplace exposure, improved access to
health care or sick leave; see \citet{baron:2012} for further discussion.

Due to its derivation from a rank-dependent social evaluation function,
as well as its origins as a measure of divergence between probability
distributions, the rank-dependent RI provides a unified mathematical
framework for modeling and/or eliciting various societal positions with
regards to public health policy. Do we favor prioritizing population
groups with lower SES (increasing $\nu>1$) because, as it may be, those
groups are more likely to utilize costly public programs? For a given
priority ranking on the SES groups and a desired health achievement
level for the population, what are the societal costs of
nonintervention? Is it realistic to expect all groups to attain the
best group rate ($\alpha\rightarrow\infty$)? Those policy-related
questions are beyond the scope of this paper. Rather, the aim of this
paper is to provide a platform that facilitates their discussion. Of
course, public programs, whether costly or not, do not only benefit
those groups with lower SES; they also benefit groups with higher SES.
Thus, the aforementioned societal costs of nonintervention are not
limited to deciding whether or not to have programs that impact those
in lower SES. Further, there are other equity arguments outside of the
cost and benefits of policies that also could be used to justify such
differential weighting as in (\ref{differential}) when measuring
socioeconomic health disparities [\citet{wilson:2009,braveman:2006}].
For instance, social justice principles remain foundational in
socioeconomic inequality measurement [\citet
{bommier:stecklov:2002,peter:2001}]. The analyst should be advised
that, while a cost-benefit
justification does not commit him/her to an ethical theory {a
priori}, cost-benefit analyses are inherently grounded in utilitarian
principles.

Even though health disparity indices are useful in that they summarize
the relationship between the distributions of disease burden and
population shares, they do not replace in-depth scientific
investigation into the complex causal pathways underlying various
health outcomes. The value of health disparity indices, such as the
slope index of inequality, the concentration index or the proposed
rank-dependent R\'{e}nyi index, is best appreciated when comparisons
between different populations as well as between different time periods
are desired, because the alternative option of tracking multiple
pairwise between-group comparisons over time can be prohibitive---as
mentioned earlier, large indicator initiatives such as HP2020 can house
over 1200 health indicators. As such, health disparity indices remain
essential for tracking the nation's progress toward the overarching
goal of achieving health equity. Like the slope and concentration
indices, as well as competing index classes, the rank-dependent RI
introduced in this paper accounts for the socioeconomic gradient in
health outcomes. However, unlike competing index classes, the
rank-dependent RI seems more stable relative to shifts in the
underlying distributions. It also allows the analyst to be explicit
about value judgment regarding the degree of societal aversion to
health inequality and the differential weighting of groups relative to
their socioeconomic rank.

%
\begin{appendix}\label{app}
\section{Derivation from weighted least squares}\label{rankG2}
Using the notation from Section~\ref{rankG} for $\nu>1$, the weighted
least-squares regression of the power-transformed outcomes $f_\alpha
(\bar{y}_{\cdot j})$ onto the standardized socioeconomic rankings
$\bar
{w}_\nu(R_j)$, with weights $p_j$, has slope and intercept parameters,
respectively,
\[
\mathrm{b}(\nu,\alpha)=\frac{S^*(\nu,\alpha)-S^*(1,\alpha)}{
{W_2(\nu)}/{W_1(\nu)^2}-1}\quad \mbox{and} \quad \mathrm{a}(\nu,\alpha )=S^*(1,
\alpha)-\mathrm{b}(\nu,\alpha),
\]
where $S^*(\nu,\alpha)$ is the (weighted) product moment between the
$f_\alpha(\bar{y}_{\cdot j})$ and $\bar{w}_\nu(R_j)$, $S^*(1,\alpha)$
is the (weighted) mean of the $f_\alpha(\bar{y}_{\cdot j})$, and the
term $\frac{W_2(\nu)}{W_1(\nu)^2} - 1$ is the (weighted) variance of
the $\bar{w}_\nu(R_j)$. (The weighted mean of the latter is 1.) From
(\ref{rank-renyi}), it follows that
\[
\operatorname{RI}_\alpha^{(\nu)}=-\ln \biggl\{\frac{f_\alpha^{-1}
[\mathrm
{a}(\nu,\alpha)+({W_2(\nu)}/{W_1(\nu)^2})\mathrm{b}(\nu,\alpha
)
]}{\mathrm{a}(\nu,0)+({W_2(\nu)}/{W_1(\nu)^2})\mathrm{b}(\nu
,0)} \biggr\}.
\]
The quantities $\mathrm{b}(\nu,0)$ and $\mathrm{b}(2,0)$ are akin to the
extended and classical slope indices of inequality, respectively [\citet
{wagstaff:2002,wagstaff:etal:1991}].

\textit{Remark}. An even more ``convenient regression'' results in the
direct interpretation of the $S^*(\nu, \alpha)$ as the slope of the
line for the regression of the following linear transform onto the
$\bar
{w}_\nu(R_j)$:
\[
\biggl(\frac{W_2(\nu)}{W_1(\nu)^2}-1 \biggr)f_\alpha(\bar{y}_{\cdot
j})+S^*(1,
\alpha)\bar{w}_\nu(R_j).
\]

\section{Sampling variability}\label{stderrors}

\subsection{NHANES data}\label{NHANESstderrors}
Total statistics are defined as follows for any scalar $a$ [\citet{talih:2012}]:
%
\begin{eqnarray}
U_{a j}&=&\sum_{s=1}^S
\sum_{c=1}^{C_s}\sum
_{i=1}^{l_{cs}}\delta _{icsj}\omega_{ics}y^a_{ics},\label{Uj}
\\
\label{Udot}
U_{a\cdot}&=&\sum_{j=1}^M
U_{a
j}.
\end{eqnarray}
In (\ref{Uj}) and (\ref{Udot}), $S$ is the number of strata; $C_s$ is
the number of PSU's in stratum~$s$; $l_{cs}$ is the number of sample
observations in the PSU-stratum pair $(c,s)$; $\omega_{ics}$ is the
sampling weight for observation $i$ in the PSU-stratum pair $(c,s)$;
$y_{ics}$ is the indicator of the adverse health outcome for
observation $i$ in the PSU-stratum pair $(c,s)$; $\delta_{icsj}=1$ when
observation $i$ [in PSU-stratum pair $(c,s)$] belongs to group $j$ and
$\delta_{icsj}=0$ otherwise; and $j$ ranges from $1$ to $M$, where $M$
is the number of groups in the population. Using the above notation, we
have $n_j/n=U_{0 j}/U_{0 \cdot}$ and $\bar{y}_{\cdot j}=U_{1 j}/U_{0
j}$. Further, define
\[
V_{0 j}=\frac{U_{0j}}{2}+\sum_{\ell=j+1}^M
U_{0\ell}.
\]
Then $(1-R_j)=V_{0j}/U_{0\cdot}$. Using these total statistics, the
rank-dependent RI in~(\ref{achieve-ratio}) is re-expressed.
For $\alpha\neq1$,
\begin{eqnarray*}
\operatorname{RI}_\alpha^{(\nu)}&=&\underbrace{\ln \Biggl[\sum
_{j=1}^M U_{1j}V_{0j}^{\nu-1}
\Biggr]}_{(\mathrm{I})}-\frac{1}{1-\alpha
}\underbrace {\ln \Biggl[\sum
_{j=1}^M U_{0j}(U_{1j}/U_{0j})^{1-\alpha}V_{0j}^{\nu
-1}
\Biggr]}_{(\mathrm{II})}\\
&&{}+\frac{\alpha}{1-\alpha}\underbrace{\ln \Biggl[\sum
_{j=1}^M U_{0j}V_{0j}^{\nu-1}
\Biggr]}_{(\mathrm{III})}.
\end{eqnarray*}
For $\alpha=1$,
\[
\operatorname{RI}_1^{(\nu)}=\ln \Biggl[\sum
_{j=1}^M U_{1j}V_{0j}^{\nu-1}
\Biggr]-\underbrace{\frac{\sum_{j=1}^M U_{0j}\ln(U_{1j}/U_{0j})
V_{0j}^{\nu
-1}}{\sum_{j=1}^M U_{0j}V_{0j}^{\nu-1}}}_{(\mathrm{IV})}-\ln \Biggl[\sum
_{j=1}^M U_{0j}V_{0j}^{\nu-1}
\Biggr].
\]

Introduce an artificial variable $\sigma_{icsk}$ that represents the
variance contribution from each sample observation. The $\sigma_{icsk}$
are obtained by taking the dot product of the vector of partial
derivatives of the rank-dependent RI with the vector of summands in the
total statistics $U_{0 k}$ and $U_{1 k}$:
%
\begin{equation}
\label{sigmas} \sigma_{icsk}= \delta_{icsk}w_{ics}
\biggl\{\frac{\partial
\operatorname
{RI}^{(\nu)}_\alpha}{\partial U_{0k}}+y_{ics}\frac{\partial
\operatorname
{RI}^{(\nu)}_\alpha}{\partial U_{1k}} \biggr\}.
\end{equation}
An estimate of the sample variance of $\operatorname{RI}^{(\nu
)}_\alpha$ is
given by the sampling variance of the total statistic $\sum_{k=1}^M\sum_{i=1}^{l_{cs}}\sigma_{icsk}$. The latter is available using
design-based estimation of variances of totals (``svytotal'') in the R
package ``survey'' [\citeauthor{lumley:2004} (\citeyear{lumley:2004,Rsurvey:2011}),
\citet{R:2011}].

\textit{Expressions for partial derivatives with respect to $U_{0
k}$ and $U_{1 k}$}.
\begin{eqnarray*}
\frac{\partial V_{0j}}{\partial U_{0k}} &=& %
\cases{0,&\quad $\mbox{if } j>k,$\vspace*{2pt}
\cr
1/2, &\quad$\mbox{if } j=k,$\vspace*{2pt}
\cr
1,&\quad $\mbox{if } j<k,$}\quad
\mbox{and}\quad
\frac{\partial V_{0j}}{\partial U_{1k}}=0,
\\
\frac{\partial}{\partial U_{0k}} (\mathrm{I}) & =&
 \frac{({(\nu-1)}/{2})U_{1k}V_{0k}^{\nu-2} + (\nu-1)\sum_{j=0}^{k-1}U_{1j}V_{0j}^{\nu
-2}}{\sum_{j=1}^M U_{1j} V_{0j}^{\nu-1}},
\\
\frac{\partial}{\partial U_{1k}} (\mathrm{I}) &=& \frac{V_{0k}^{\nu
-1}}{\sum_{j=1}^M U_{1j} V_{0j}^{\nu-1}},
\\
\frac{\partial}{\partial U_{0k}} (\mathrm{II}) & =& \Biggl(\alpha
(U_{1k}/U_{0k})^{1-\alpha}V_{0k}^{\nu-1} + \frac{\nu
-1}{2}U_{0k}(U_{1k}/U_{0k})^{1-\alpha}V_{0k}^{\nu-2} \\
&&\hspace*{78pt}{}+
(\nu-1)\sum_{j=0}^{k-1}U_{0j}(U_{1j}/U_{0j})^{1-\alpha}V_{0j}^{\nu-2}\Biggr)\\
&&{}\bigg/\Biggl(\sum_{j=1}^M U_{0j}(U_{1j}/U_{0j})^{1-\alpha} V_{0j}^{\nu-1}\Biggr),
\\
\frac{\partial}{\partial U_{1k}} (\mathrm{II}) & =& \frac{(1-\alpha
)(U_{1k}/U_{0k})^{-\alpha}V_{0k}^{\nu-1}}{\sum_{j=1}^M
U_{0j}(U_{1j}/U_{0j})^{1-\alpha} V_{0j}^{\nu-1}},
\\
\frac{\partial}{\partial U_{0k}} (\mathrm{III}) & =& \frac{V_{0k}^{\nu
-1} + ({(\nu-1)}/{2})U_{0k}V_{0k}^{\nu-2} + (\nu-1)\sum_{j=0}^{k-1}U_{0j}V_{0j}^{\nu-2}}{\sum_{j=1}^M U_{0j} V_{0j}^{\nu-1}},
\\
\frac{\partial}{\partial U_{1k}} (\mathrm{III}) &=& 0,
\\
\frac{\partial}{\partial U_{0k}} (\mathrm{IV}) & =& \Biggl(\bigl[\ln
(U_{1k}/U_{0k}) - 1\bigr] V_{0k}^{\nu-1} + \frac{\nu-1}{2}U_{0k}\ln
(U_{1k}/U_{0k})V_{0k}^{\nu-2} \\
&&\hspace*{89pt}{}+ (\nu-1)\sum_{j=0}^{k-1}U_{0j}\ln
(U_{1j}/U_{0j})V_{0j}^{\nu-2}\Biggr)\\
&&{}\bigg/\Biggl(\sum_{j=1}^M U_{0j}V_{0j}^{\nu-1}\Biggr)
\\
&&{} - \Biggl(\Biggl [V_{0k}^{\nu-1} + \frac{\nu
-1}{2}U_{0k}V_{0k}^{\nu-2}\\
&&\hspace*{25pt}{} + (\nu-1)\sum_{j=0}^{k-1}U_{0j}V_{0j}^{\nu
-2} \Biggr]  \Biggl[ \sum_{j=1}^M U_{0j} \ln(U_{1j}/U_{0j})V_{0j}^{\nu-1}
 \Biggr]\Biggr)\\
 &&{}\bigg/\Biggl (\sum_{j=1}^M U_{0j} V_{0j}^{\nu-1} \Biggr)^2,
\\
\frac{\partial}{\partial U_{1k}} (\mathrm{IV}) &=& \frac
{(U_{1k}/U_{0k})^{-1}V_{0k}^{\nu-1}}{\sum_{j=1}^M U_{0j} V_{0j}^{\nu-1}}.
\end{eqnarray*}

\textit{Notes}. NHANES has a stratified multistage probability
sampling design structure [\citet{NHANESguidelines:2013}]. While the
sample weights provided in the NHANES public-use data files reflect the
unequal probabilities of selection, they also reflect nonresponse
adjustments and adjustments to independent population controls.
Therefore, strictly speaking, they are not the true sampling weights
$w_{ics}$ in~(\ref{Uj}).

\subsection{SEER data}\label{SEERstderrors}
Following SEER*Stat [\citet{SEERSoft:2013}], crude rates are assumed to
be distributed according to a Poisson distribution. In addition,
age-adjusted rates are adjusted using the year 2000 U.S. standard
population, with known age-adjustment weights $\omega_k$ and sizes
$n_{kj}$. Thus, sample means and variances for the age-adjusted rates
are as follows:
\[
\hat{\E}[\bar{y}_{\cdot j}]=\sum_{k=1}^K
\omega_k\bar{u}_{\cdot k j} \quad\mbox{and}\quad \widehat{\Var}[
\bar{y}_{\cdot j}]=\sum_{k=1}^K\omega
_k^2\bar {u}_{\cdot k j}/n_{kj},
\]
where the $\bar{u}_{\cdot k j}$ are the underlying crude rates for age
group k. Using the expression in (\ref{achieve-ratio}), we have
\[
\frac{\partial\operatorname{RI}_\alpha^{(\nu)}}{\partial\bar
{y}_{\cdot j}}=\bar {w}_\nu(R_j)p_j
\biggl[\frac{1}{H^*(\nu,0)}-\frac{1}{\bar
{y}_{\cdot
j}^\alpha H^*(\nu,\alpha)^{1-\alpha}} \biggr].
\]
The Taylor series linearization approximation to the variance of the
rank-dependent RI yields
\[
\widehat{\Var} \bigl[\operatorname{RI}_\alpha^{(\nu)} \bigr]=\sum
_{j=1}^M \biggl[\frac{\partial\operatorname{RI}_\alpha^{(\nu)}}{\partial\bar
{y}_{\cdot
j}}
\biggr]^2\widehat{\Var}[\bar{y}_{\cdot j}].
\]
\end{appendix}

\section*{Acknowledgments} All data were compiled from public-use files
and analyzed by the author. Thanks to Rebecca Hines, Van Parsons and
Jennifer Madans (NCHS) for their constructive feedback. Hallway
conversations with NCHS researchers Lauren Rossen, Frederic Selck and
Sirin Yaemsiri, as well as insightful comments from the journal editor
and reviewers, improved the presentation of findings. NCHS analysts
David Huang and Kimberly Hurvitz answered questions relating to HP2020
objectives C-10 and NWS-4.1, respectively.

\section*{Disclaimer}
The findings and conclusions in this paper are those of the author and
do not necessarily represent the views of the CDC or NCHS.



\begin{thebibliography}{84}

\bibitem[\protect\citeauthoryear{Aaberge}{2005}]{aaberge:2005}
%
\begin{bmisc}[author]
\bauthor{\bsnm{Aaberge},~\bfnm{Rolf}\binits{R.}}
(\byear{2005}).
\bhowpublished{Asymptotic distribution theory of empirical
rank-dependent measures of inequality.
Discussion Papers No. 402, Statistics Norway, Research Department.
Available at \surl{http://www.ssb.no/a/publikasjoner/pdf/DP/dp402.pdf}.}
\end{bmisc}
%
\bptok{imsref}%
\endbibitem

\bibitem[\protect\citeauthoryear{Akers, Newman and
Smith}{2007}]{akers:etal:2007}
%
\begin{barticle}[author]
\bauthor{\bsnm{Akers},~\bfnm{A.~Y.}\binits{A.~Y.}},
\bauthor{\bsnm{Newman},~\bfnm{S.~J.}\binits{S.~J.}} \AND
\bauthor{\bsnm{Smith},~\bfnm{J.~S.}\binits{J.~S.}}
(\byear{2007}).
\btitle{Factors underlying disparities in cervical cancer incidence,
screening, and treatment in the United States}.
\bjournal{Curr. Probl. Cancer}
\bvolume{31}
\bpages{157--181}.
\end{barticle}
%
\bptok{imsref}%
\endbibitem

\bibitem[\protect\citeauthoryear{Asada, Yoshida and
Whipp}{2013}]{asada:etal:2013}
%
\begin{barticle}[pbm]
\bauthor{\bsnm{Asada},~\bfnm{Yukiko}\binits{Y.}},
\bauthor{\bsnm{Yoshida},~\bfnm{Yoko}\binits{Y.}} \AND
\bauthor{\bsnm{Whipp},~\bfnm{Alyce~M.}\binits{A.~M.}}
(\byear{2013}).
\btitle{Summarizing social disparities in health}.
\bjournal{Milbank Q.}
\bvolume{91}
\bpages{5--36}.
\bid{doi={10.1111/milq.12001}, issn={1468-0009}, pmcid={3607125},
pmid={23488710}}
\end{barticle}
%
\bptok{imsref}%
\endbibitem

\bibitem[\protect\citeauthoryear{Atkinson}{1970}]{atkinson:1970}
%
\begin{barticle}[mr]
\bauthor{\bsnm{Atkinson},~\bfnm{Anthony~B.}\binits{A.~B.}}
(\byear{1970}).
\btitle{On the measurement of inequality}.
\bjournal{J. Econom. Theory}
\bvolume{2}
\bpages{244--263}.
\bid{issn={0022-0531}, mr={0449508}}
\end{barticle}
%
\bptok{imsref}%
\endbibitem

\bibitem[\protect\citeauthoryear{Baron}{2012}]{baron:2012}
%
\begin{bmisc}[author]
\bauthor{\bsnm{Baron},~\bfnm{Sherry}\binits{S.}}
(\byear{2012}).
\bhowpublished{Measuring occupation as an element of
socioeconomic status/position.
Presented at the March 2012 Hearing of the National
Committee on Vital and Health Statistics on Minimum
Data Standards for the Measurement of Socioeconomic
Status in Federal Health Surveys.
Available at \surl{http://www.ncvhs.hhs.gov/120308p2.pdf}.}
\end{bmisc}
%
\bptok{imsref}%
\endbibitem

\bibitem[\protect\citeauthoryear{Bennett and
Mitra}{2013}]{bennett:mitra:2013}
%
\begin{barticle}[mr]
\bauthor{\bsnm{Bennett},~\bfnm{Christopher~J.}\binits{C.~J.}} \AND
\bauthor{\bsnm{Mitra},~\bfnm{Shabana}\binits{S.}}
(\byear{2013}).
\btitle{Multidimensional poverty: Measurement, estimation, and inference}.
\bjournal{Econometric Rev.}
\bvolume{32}
\bpages{57--83}.
\bid{doi={10.1080/07474938.2012.690331}, issn={0747-4938}, mr={2988920}}
\end{barticle}
%
\bptok{imsref}%
\endbibitem

\bibitem[\protect\citeauthoryear{Berrebi and
Silber}{1981}]{berrebi:silber:1981}
%
\begin{barticle}[author]
\bauthor{\bsnm{Berrebi},~\bfnm{Z.~M.}\binits{Z.~M.}} \AND
\bauthor{\bsnm{Silber},~\bfnm{Jacques}\binits{J.}}
(\byear{1981}).
\btitle{Weighting income ranks and levels: A multiple-parameter
generalization for absolute and relative inequality indices}.
\bjournal{Econom. Lett.}
\bvolume{7}
\bpages{391--397}.
\end{barticle}
%
\bptok{imsref}%
\endbibitem

\bibitem[\protect\citeauthoryear{Biewen and
Jenkins}{2006}]{biewen:jenkins:2006}
%
\begin{barticle}[author]
\bauthor{\bsnm{Biewen},~\bfnm{Martin}\binits{M.}} \AND
\bauthor{\bsnm{Jenkins},~\bfnm{Stephen~P.}\binits{S.~P.}}
(\byear{2006}).
\btitle{Variance estimation for generalized entropy and Atkinson
inequality indices: The complex survey data case}.
\bjournal{Oxford Bulletin of Economics and Statistics}
\bvolume{68}
\bpages{371--383}.
\end{barticle}
%
\bptok{imsref}%
\endbibitem

\bibitem[\protect\citeauthoryear{Bleichrodt, Rohde and
Ourti}{2012}]{bleichrodt:etal:2012}
%
\begin{barticle}[pbm]
\bauthor{\bsnm{Bleichrodt},~\bfnm{Han}\binits{H.}},
\bauthor{\bsnm{Rohde},~\bfnm{Kirsten~I.~M.}\binits{K.~I.~M.}} \AND
\bauthor{\bsnm{Van Ourti},~\bfnm{Tom}\binits{T.}}
(\byear{2012}).
\btitle{An experimental test of the concentration index}.
\bjournal{J. Health Econ.}
\bvolume{31}
\bpages{86--98}.
\bid{doi={10.1016/j.jhealeco.2011.12.003}, issn={1879-1646},
pii={S0167-6296(11)00169-X}, pmid={22307035}}
\end{barticle}
%
\bptok{imsref}%
\endbibitem

\bibitem[\protect\citeauthoryear{Bleichrodt and van
Doorslaer}{2006}]{bleichrodt:vanDoorslaer:2006}
%
\begin{barticle}[pbm]
\bauthor{\bsnm{Bleichrodt},~\bfnm{Han}\binits{H.}} \AND
\bauthor{\bparticle{van} \bsnm{Doorslaer},~\bfnm{Eddy}\binits{E.}}
(\byear{2006}).
\btitle{A welfare economics foundation for health inequality measurement}.
\bjournal{J. Health Econ.}
\bvolume{25}
\bpages{945--957}.
\bid{doi={10.1016/j.jhealeco.2006.01.002}, issn={0167-6296},
pii={S0167-6296(06)00003-8}, pmid={16466818}}
\end{barticle}
%
\bptok{imsref}%
\endbibitem

\bibitem[\protect\citeauthoryear{Bommier and
Stecklov}{2002}]{bommier:stecklov:2002}
%
\begin{barticle}[pbm]
\bauthor{\bsnm{Bommier},~\bfnm{Antoine}\binits{A.}} \AND
\bauthor{\bsnm{Stecklov},~\bfnm{Guy}\binits{G.}}
(\byear{2002}).
\btitle{Defining health inequality: Why Rawls succeeds where social
welfare theory fails}.
\bjournal{J. Health Econ.}
\bvolume{21}
\bpages{497--513}.
\bid{issn={0167-6296}, pii={S0167-6296(01)00138-2}, pmid={12022270}}
\end{barticle}
%
\bptok{imsref}%
\endbibitem

\bibitem[\protect\citeauthoryear{Borrell and Talih}{2012}]{borrell:talih:2012}
%
\begin{barticle}[pbm]
\bauthor{\bsnm{Borrell},~\bfnm{Luisa~N.}\binits{L.~N.}} \AND
\bauthor{\bsnm{Talih},~\bfnm{Makram}\binits{M.}}
(\byear{2012}).
\btitle{Examining periodontal disease disparities among U.S. adults 20
years of age and older: NHANES III (1988--1994) and NHANES 1999--2004}.
\bjournal{Public Health Rep.}
\bvolume{127}
\bpages{497--506}.
\bid{issn={1468-2877}, pmcid={3407849}, pmid={22942467}}
\end{barticle}
%
\bptok{imsref}%
\endbibitem

\bibitem[\protect\citeauthoryear{Borrell and
Talih}{2011}]{borrell:talih:2011}
%
\begin{barticle}[mr]
\bauthor{\bsnm{Borrell},~\bfnm{Luisa~N.}\binits{L.~N.}} \AND
\bauthor{\bsnm{Talih},~\bfnm{Makram}\binits{M.}}
(\byear{2011}).
\btitle{A symmetrized {T}heil index measure of health disparities: An
example using dental caries in U.{S}. children and adolescents}.
\bjournal{Stat. Med.}
\bvolume{30}
\bpages{277--290}.
\bid{doi={10.1002/sim.4114}, issn={0277-6715}, mr={2758878}}
\end{barticle}
%
\bptok{imsref}%
\endbibitem

\bibitem[\protect\citeauthoryear{Boyd and Nam}{2004}]{boyd:nam:2004}
%
\begin{barticle}[author]
\bauthor{\bsnm{Boyd},~\bfnm{Monica}\binits{M.}} \AND
\bauthor{\bsnm{Nam},~\bfnm{Charles~B.}\binits{C.~B.}}
(\byear{2004}).
\btitle{Occupational status in 2000: Over a century of census-based
measurement}.
\bjournal{Popul. Res. Policy Rev.}
\bvolume{23}
\bpages{327--358}.
\end{barticle}
%
\bptok{imsref}%
\endbibitem

\bibitem[\protect\citeauthoryear{Braveman}{2006}]{braveman:2006}
%
\begin{barticle}[pbm]
\bauthor{\bsnm{Braveman},~\bfnm{Paula}\binits{P.}}
(\byear{2006}).
\btitle{Health disparities and health equity: Concepts and measurement}.
\bjournal{Annu. Rev. Public Health}
\bvolume{27}
\bpages{167--194}.
\bid{doi={10.1146/annurev.publhealth.27.021405.102103},
issn={0163-7525}, pmid={16533114}}
\end{barticle}
%
\bptok{imsref}%
\endbibitem

\bibitem[\protect\citeauthoryear{Braveman et~al.}{2010}]{braveman:etal:2010}
%
\begin{barticle}[pbm]
\bauthor{\bsnm{Braveman},~\bfnm{Paula~A.}\binits{P.~A.}},
\bauthor{\bsnm{Cubbin},~\bfnm{Catherine}\binits{C.}},
\bauthor{\bsnm{Egerter},~\bfnm{Susan}\binits{S.}},
\bauthor{\bsnm{Williams},~\bfnm{David~R.}\binits{D.~R.}} \AND
\bauthor{\bsnm{Pamuk},~\bfnm{Elsie}\binits{E.}}
(\byear{2010}).
\btitle{Socioeconomic disparities in health in the United States: What
the patterns tell us}.
\bjournal{Am. J. Public Health}
\bvolume{100}
\bpages{S186--S196}.
\bid{doi={10.2105/AJPH.2009.166082}, issn={1541-0048},
pii={AJPH.2009.166082}, pmcid={2837459}, pmid={20147693}}
\bptnote{check volume}%
\end{barticle}
%
\bptok{imsref}%
\endbibitem

\bibitem[\protect\citeauthoryear{Chen, Roy and
Crawford}{2012}]{chen:etal:2012}
%
\begin{barticle}[pbm]
\bauthor{\bsnm{Chen},~\bfnm{Zhuo}\binits{Z.}},
\bauthor{\bsnm{Roy},~\bfnm{Kakoli}\binits{K.}} \AND
\bauthor{\bsnm{Crawford},~\bfnm{Carol~A.~Gotway}\binits{C.~A.~G.}}
(\byear{2012}).
\btitle{Evaluation of variance estimators for the concentration and
health achievement indices: A Monte Carlo simulation}.
\bjournal{Health Econ.}
\bvolume{21}
\bpages{1375--1381}.
\bid{doi={10.1002/hec.1796}, issn={1099-1050}, pmid={21956946}}
\end{barticle}
%
\bptok{imsref}%
\endbibitem

\bibitem[\protect\citeauthoryear{Chen et~al.}{2013}]{chen:etal:2013}
%
\begin{barticle}[pbm]
\bauthor{\bsnm{Chen},~\bfnm{Jarvis~T.}\binits{J.~T.}},
\bauthor{\bsnm{Beckfield},~\bfnm{Jason}\binits{J.}},
\bauthor{\bsnm{Waterman},~\bfnm{Pamela~D.}\binits{P.~D.}} \AND
\bauthor{\bsnm{Krieger},~\bfnm{Nancy}\binits{N.}}
(\byear{2013}).
\btitle{Can changes in the distributions of and associations between
education and income bias temporal comparisons of health disparities?
An exploration with causal graphs and simulations}.
\bjournal{Am. J. Epidemiol.}
\bvolume{177}
\bpages{870--881}.
\bid{doi={10.1093/aje/kwt041}, issn={1476-6256}, pii={kwt041},
pmcid={4023297}, pmid={23568593}}
\end{barticle}
%
\bptok{imsref}%
\endbibitem

\bibitem[\protect\citeauthoryear{Cheng, Han and
Gansky}{2008}]{cheng:etal:2008}
%
\begin{barticle}[pbm]
\bauthor{\bsnm{Cheng},~\bfnm{Nancy~F.}\binits{N.~F.}},
\bauthor{\bsnm{Han},~\bfnm{Pamela~Z.}\binits{P.~Z.}} \AND
\bauthor{\bsnm{Gansky},~\bfnm{Stuart~A.}\binits{S.~A.}}
(\byear{2008}).
\btitle{Methods and software for estimating health disparities: The
case of children's oral health}.
\bjournal{Am. J. Epidemiol.}
\bvolume{168}
\bpages{906--914}.
\bid{doi={10.1093/aje/kwn207}, issn={1476-6256}, mid={NIHMS63747},
pii={kwn207}, pmcid={2597673}, pmid={18779387}}
\end{barticle}
%
\bptok{imsref}%
\endbibitem

\bibitem[\protect\citeauthoryear{Chernoff}{1952}]{chernoff:1952}
%
\begin{barticle}[mr]
\bauthor{\bsnm{Chernoff},~\bfnm{Herman}\binits{H.}}
(\byear{1952}).
\btitle{A measure of asymptotic efficiency for tests of a hypothesis
based on the sum of observations}.
\bjournal{Ann. Math. Statistics}
\bvolume{23}
\bpages{493--507}.
\bid{issn={0003-4851}, mr={0057518}}
\end{barticle}
%
\bptok{imsref}%
\endbibitem

\bibitem[\protect\citeauthoryear{Costa-Font and Hern{\'
{a}}ndez-Quevedo}{2012}]{costa:quevedo:2012}
%
\begin{barticle}[pbm]
\bauthor{\bsnm{Costa-Font},~\bfnm{Joan}\binits{J.}} \AND
\bauthor{\bsnm{Hern{\'{a}}ndez-Quevedo},~\bfnm{Cristina}\binits{C.}}
(\byear{2012}).
\btitle{Measuring inequalities in health: What do we know? What do we
need to know?}
\bjournal{Health Policy}
\bvolume{106}
\bpages{195--206}.
\bid{doi={10.1016/j.healthpol.2012.04.007}, issn={1872-6054},
pii={S0168-8510(12)00112-1}, pmid={22607941}}
\end{barticle}
%
\bptok{imsref}%
\endbibitem

\bibitem[\protect\citeauthoryear{Cowell, Davidson and
Flachaire}{2011}]{cowell:etal:2011}
%
\begin{bmisc}[author]
\bauthor{\bsnm{Cowell},~\bfnm{Frank~A.}\binits{F.~A.}},
\bauthor{\bsnm{Davidson},~\bfnm{Russel}\binits{R.}} \AND
\bauthor{\bsnm{Flachaire},~\bfnm{Emmanuel}\binits{E.}}
(\byear{2011}).
\bhowpublished{Goodness of fit: An axiomatic approach.
Discussion Paper DT 2011-50, Groupement de Recherche en Economie
Quantitative d'Aix-Marseille (GREQAM).
Available at \surl{http://halshs.\\archives-ouvertes.fr/docs/00/63/90/75/PDF/DTGREQAM2011\_50.pdf}}.
\end{bmisc}
%
\bptok{imsref}%
\endbibitem

\bibitem[\protect\citeauthoryear{Cowell and
Gardiner}{1999}]{cowell:gardiner:1999}
%
\begin{bmisc}[author]
\bauthor{\bsnm{Cowell},~\bfnm{Frank~A.}\binits{F.~A.}} \AND
\bauthor{\bsnm{Gardiner},~\bfnm{Karen}\binits{K.}}
(\byear{1999}).
\bhowpublished{Welfare weights.
OFT Research Paper 202,
STICERD---London School of Economics. Available at\newline
\href{http://darp.lse.\\ac.uk/papersDB/Cowell-Gardiner\_\%28OFT\%29.pdf}
{http://darp.lse.ac.uk/papersDB/Cowell-Gardiner\_(OFT).pdf}.}
\end{bmisc}
%
\bptok{imsref}%
\endbibitem

\bibitem[\protect\citeauthoryear{Cowell and Kuga}{1981}]{cowell:kuga:1981}
%
\begin{barticle}[mr]
\bauthor{\bsnm{Cowell},~\bfnm{Frank~A.}\binits{F.~A.}} \AND
\bauthor{\bsnm{Kuga},~\bfnm{Kiyoshi}\binits{K.}}
(\byear{1981}).
\btitle{Additivity and the entropy concept: An axiomatic approach to
inequality measurement}.
\bjournal{J. Econom. Theory}
\bvolume{25}
\bpages{131--143}.
\bid{doi={10.1016/0022-0531(81)90020-X}, issn={0022-0531}, mr={0636017}}
\end{barticle}
%
\bptok{imsref}%
\endbibitem

\bibitem[\protect\citeauthoryear{Dean, Williams and
Fenton}{2013}]{dean:etal:2013}
%
\begin{barticle}[pbm]
\bauthor{\bsnm{Dean},~\bfnm{Hazel~D.}\binits{H.~D.}},
\bauthor{\bsnm{Williams},~\bfnm{Kim~M.}\binits{K.~M.}} \AND
\bauthor{\bsnm{Fenton},~\bfnm{Kevin~A.}\binits{K.~A.}}
(\byear{2013}).
\btitle{From theory to action: Applying social determinants of health
to public health practice}.
\bjournal{Public Health Rep.}
\bvolume{128}
\bpages{S31--S34}.
\bid{issn={1468-2877}, pmcid={3945442}, pmid={24179272}}
\bptnote{check volume}%
\end{barticle}
%
\bptok{imsref}%
\endbibitem

\bibitem[\protect\citeauthoryear{Decancq and Lugo}{2009}]{decancq:lugo:2009}
%
\begin{bmisc}[author]
\bauthor{\bsnm{Decancq},~\bfnm{Koen}\binits{K.}} \AND
\bauthor{\bsnm{Lugo},~\bfnm{Mar\'ia~Ana}\binits{M.~A.}}
(\byear{2009}).
\bhowpublished{Measuring inequality of well-being with a
correlation-sensitive multidimensional Gini index.
Working paper ECINEQ 2009-124, Society for the Study of Economic Inequality.
Available at \surl{http://www.ecineq.org/\\milano/WP/ECINEQ2009-124.pdf}.}
\end{bmisc}
%
\bptok{imsref}%
\endbibitem

\bibitem[\protect\citeauthoryear{{DHHS}}{2014}]{HP.gov:2013}
%
\begin{bmisc}[author]
\borganization{DHHS}
(\byear{2014}).
\bhowpublished{HealthyPeople.gov.
U.S. Department of Health and Human Services (DHHS),
Washington, DC.
Available at \surl{http://healthypeople.gov}}.
\end{bmisc}
%
\bptok{imsref}%
\endbibitem

\bibitem[\protect\citeauthoryear{Erreygers}{2009a}]{erreygers:2009shortfall}
%
\begin{barticle}[pbm]
\bauthor{\bsnm{Erreygers},~\bfnm{Guido}\binits{G.}}
(\byear{2009}a).
\btitle{Can a single indicator measure both attainment and shortfall
inequality?}
\bjournal{J. Health Econ.}
\bvolume{28}
\bpages{885--893}.
\bid{doi={10.1016/j.jhealeco.2009.03.005}, issn={0167-6296},
pii={S0167-6296(09)00038-1}, pmid={19409631}}
\end{barticle}
%
\bptok{imsref}%
\endbibitem

\bibitem[\protect\citeauthoryear{Erreygers}{2009b}]{erreygers:2009correcting}
%
\begin{barticle}[pbm]
\bauthor{\bsnm{Erreygers},~\bfnm{Guido}\binits{G.}}
(\byear{2009}b).
\btitle{Correcting the concentration index}.
\bjournal{J. Health Econ.}
\bvolume{28}
\bpages{504--515}.
\bid{doi={10.1016/j.jhealeco.2008.02.003}, issn={0167-6296},
pii={S0167-6296(08)00007-6}, pmid={18367273}}
\bptnote{check related}%
\end{barticle}
%
\bptok{imsref}%
\endbibitem

\bibitem[\protect\citeauthoryear{Erreygers and {van
Ourti}}{2011}]{erreygers:vanOurti:2011}
%
\begin{barticle}[author]
\bauthor{\bsnm{Erreygers},~\bfnm{Guido}\binits{G.}} \AND
\bauthor{\bsnm{{van Ourti}},~\bfnm{Tom}\binits{T.}}
(\byear{2011}).
\btitle{Measuring socioeconomic inequality in health, health care and
health financing by mean of rank-dependent indices: A recipe for good practice}.
\bjournal{J. Health Econ.}
\bvolume{30}
\bpages{685--694}.
\end{barticle}
%
\bptok{imsref}%
\endbibitem

\bibitem[\protect\citeauthoryear{Foster, McGillivray and
Seth}{2013}]{foster:etal:2013}
%
\begin{barticle}[mr]
\bauthor{\bsnm{Foster},~\bfnm{James~E.}\binits{J.~E.}},
\bauthor{\bsnm{McGillivray},~\bfnm{Mark}\binits{M.}} \AND
\bauthor{\bsnm{Seth},~\bfnm{Suman}\binits{S.}}
(\byear{2013}).
\btitle{Composite indices: Rank robustness, statistical association,
and redundancy}.
\bjournal{Econometric Rev.}
\bvolume{32}
\bpages{35--56}.
\bid{doi={10.1080/07474938.2012.690647}, issn={0747-4938}, mr={2988919}}
\end{barticle}
%
\bptok{imsref}%
\endbibitem

\bibitem[\protect\citeauthoryear{Frohlich and
Potvin}{2008}]{frohlich:potvin:2008}
%
\begin{barticle}[pbm]
\bauthor{\bsnm{Frohlich},~\bfnm{Katherine~L.}\binits{K.~L.}} \AND
\bauthor{\bsnm{Potvin},~\bfnm{Louise}\binits{L.}}
(\byear{2008}).
\btitle{The
inequality paradox: The population approach and vulnerable populations}.
\bjournal{Am. J. Public Health}
\bvolume{98}
\bpages{216--221}.
\bid{doi={10.2105/AJPH.2007.114777}, issn={1541-0048},
pii={AJPH.2007.114777}, pmcid={2376882}, pmid={18172133}}
\end{barticle}
%
\bptok{imsref}%
\endbibitem

\bibitem[\protect\citeauthoryear{Gravelle}{2003}]{gravelle:2003}
%
\begin{barticle}[pbm]
\bauthor{\bsnm{Gravelle},~\bfnm{Hugh}\binits{H.}}
(\byear{2003}).
\btitle{Measuring income related inequality in health: Standardisation
and the partial concentration index}.
\bjournal{Health Econ.}
\bvolume{12}
\bpages{803--819}.
\bid{doi={10.1002/hec.813}, issn={1057-9230}, pmid={14508866}}
\end{barticle}
%
\bptok{imsref}%
\endbibitem

\bibitem[\protect\citeauthoryear{Harper et~al.}{2008}]{harper:etal:2008}
%
\begin{barticle}[pbm]
\bauthor{\bsnm{Harper},~\bfnm{Sam}\binits{S.}},
\bauthor{\bsnm{Lynch},~\bfnm{John}\binits{J.}},
\bauthor{\bsnm{Meersman},~\bfnm{Stephen~C.}\binits{S.~C.}},
\bauthor{\bsnm{Breen},~\bfnm{Nancy}\binits{N.}},
\bauthor{\bsnm{Davis},~\bfnm{William~W.}\binits{W.~W.}} \AND
\bauthor{\bsnm{Reichman},~\bfnm{Marsha~E.}\binits{M.~E.}}
(\byear{2008}).
\btitle{An overview of methods for monitoring social disparities in
cancer with an example using trends in lung cancer incidence by
area-socioeconomic position and race-ethnicity, 1992--2004}.
\bjournal{Am. J. Epidemiol.}
\bvolume{167}
\bpages{889--899}.
\bid{doi={10.1093/aje/kwn016}, issn={1476-6256}, mid={NIHMS48152},
pii={kwn016}, pmcid={2409988}, pmid={18344513}}
\end{barticle}
%
\bptok{imsref}%
\endbibitem

\bibitem[\protect\citeauthoryear{Harper et~al.}{2010}]{harper:etal:2010}
%
\begin{barticle}[author]
\bauthor{\bsnm{Harper},~\bfnm{Sam}\binits{S.}},
\bauthor{\bsnm{King},~\bfnm{Nicholas~B.}\binits{N.~B.}},
\bauthor{\bsnm{Meersman},~\bfnm{Stephen~C.}\binits{S.~C.}},
\bauthor{\bsnm{Reichman},~\bfnm{Marsha~E.}\binits{M.~E.}},
\bauthor{\bsnm{Breen},~\bfnm{Nancy}\binits{N.}} \AND
\bauthor{\bsnm{Lynch},~\bfnm{John}\binits{J.}}
(\byear{2010}).
\btitle{Implicit value judgments in the measurement of health inequalities}.
\bjournal{Milbank Q.}
\bvolume{88}
\bpages{4--29}.
\end{barticle}
%
\bptok{imsref}%
\endbibitem

\bibitem[\protect\citeauthoryear{Johnson
et~al.}{2013}]{NHANESguidelines:2013}
%
\begin{barticle}[author]
\bauthor{\bsnm{Johnson},~\bfnm{Clifford~L.}\binits{C.~L.}},
\bauthor{\bsnm{{Paulose-Ram}},~\bfnm{Ryne}\binits{R.}},
\bauthor{\bsnm{Ogden},~\bfnm{Cynthia~L.}\binits{C.~L.}},
\bauthor{\bsnm{Carroll},~\bfnm{Margaret~D.}\binits{M.~D.}},
\bauthor{\bsnm{{Kruszan-Moran}},~\bfnm{Deanna}\binits{D.}},
\bauthor{\bsnm{Dohrmann},~\bfnm{Sylvia~M.}\binits{S.~M.}} \AND
\bauthor{\bsnm{Curtin},~\bfnm{Lester~R.}\binits{L.~R.}}
(\byear{2013}).
\btitle{National health and nutrition examination survey: Analytic
guidelines, 1999--2010}.
\bjournal{Vital and Health Statistics, Series 2}
\bvolume{161}.
\end{barticle}
%
\bptok{imsref}%
\endbibitem

\bibitem[\protect\citeauthoryear{Kakwani, Wagstaff and {van
Doorslaer}}{1997}]{kakwani:etal:1997}
%
\begin{barticle}[author]
\bauthor{\bsnm{Kakwani},~\bfnm{Nanak}\binits{N.}},
\bauthor{\bsnm{Wagstaff},~\bfnm{Adam}\binits{A.}} \AND
\bauthor{\bsnm{{van Doorslaer}},~\bfnm{Eddy}\binits{E.}}
(\byear{1997}).
\btitle{Socioeconomic inequalities in health: Measurement,
computation, and statistical inference}.
\bjournal{J. Econom.}
\bvolume{77}
\bpages{87--103}.
\end{barticle}
%
\bptok{imsref}%
\endbibitem

\bibitem[\protect\citeauthoryear{Keppel
et~al.}{2005}]{keppel:pamuk:lynch:etal:2005}
%
\begin{barticle}[author]
\bauthor{\bsnm{Keppel},~\bfnm{Kenneth}\binits{K.}},
\bauthor{\bsnm{Pamuk},~\bfnm{Elsie}\binits{E.}},
\bauthor{\bsnm{Lynch},~\bfnm{John}\binits{J.}},
\bauthor{\bsnm{{Carter-Pokras}},~\bfnm{Olivia}\binits{O.}},
\bauthor{\bsnm{Kim},~\bfnm{Insun}\binits{I.}},
\bauthor{\bsnm{Mays},~\bfnm{Vickie}\binits{V.}},
\bauthor{\bsnm{Pearcy},~\bfnm{Jeffrey}\binits{J.}},
\bauthor{\bsnm{Schoenbach},~\bfnm{Victor}\binits{V.}} \AND
\bauthor{\bsnm{Weissman},~\bfnm{Joel~S.}\binits{J.~S.}}
(\byear{2005}).
\btitle{Methodological issues in measuring health disparities}.
\bjournal{Vital and Health Statistics, Series 2}
\bvolume{141}.
\end{barticle}
%
\bptok{imsref}%
\endbibitem

\bibitem[\protect\citeauthoryear{Kjellsson and
Gerdtham}{2013}]{kjellsson:gerdtham:2013}
%
\begin{barticle}[pbm]
\bauthor{\bsnm{Kjellsson},~\bfnm{Gustav}\binits{G.}} \AND
\bauthor{\bsnm{Gerdtham},~\bfnm{Ulf-G.}\binits{U.-G.}}
(\byear{2013}).
\btitle{On correcting the concentration index for binary variables}.
\bjournal{J. Health Econ.}
\bvolume{32}
\bpages{659--670}.
\bid{doi={10.1016/j.jhealeco.2012.10.012}, issn={1879-1646},
pii={S0167-6296(12)00173-7}, pmid={23522656}}
\end{barticle}
%
\bptok{imsref}%
\endbibitem

\bibitem[\protect\citeauthoryear{Koolman and van
Doorslaer}{2004}]{koolman:vanDoorslaer:2004}
%
\begin{barticle}[pbm]
\bauthor{\bsnm{Koolman},~\bfnm{Xander}\binits{X.}} \AND
\bauthor{\bparticle{van} \bsnm{Doorslaer},~\bfnm{Eddy}\binits{E.}}
(\byear{2004}).
\btitle{On the interpretation of a concentration index of inequality}.
\bjournal{Health Econ.}
\bvolume{13}
\bpages{649--656}.
\bid{doi={10.1002/hec.884}, issn={1057-9230}, pmid={15259044}}
\end{barticle}
%
\bptok{imsref}%
\endbibitem

\bibitem[\protect\citeauthoryear{Krieger, Williams and
Moss}{1997}]{krieger:etal:1997}
%
\begin{barticle}[author]
\bauthor{\bsnm{Krieger},~\bfnm{Nancy}\binits{N.}},
\bauthor{\bsnm{Williams},~\bfnm{D.~R.}\binits{D.~R.}} \AND
\bauthor{\bsnm{Moss},~\bfnm{N.~E.}\binits{N.~E.}}
(\byear{1997}).
\btitle{Measuring social class in US public health research: Concepts,
methodologies, and guidelines}.
\bjournal{Annu. Rev. Public Health}
\bvolume{18}
\bpages{341--378}.
\end{barticle}
%
\bptok{imsref}%
\endbibitem

\bibitem[\protect\citeauthoryear{Kuczmarski et~al.}{2002}]{growthcharts:2002}
%
\begin{barticle}[author]
\bauthor{\bsnm{Kuczmarski},~\bfnm{Robert~J.}\binits{R.~J.}},
\bauthor{\bsnm{Ogden},~\bfnm{Cynthia~L.}\binits{C.~L.}},
\bauthor{\bsnm{Guo},~\bfnm{Shumei~S.}\binits{S.~S.}},
\bauthor{\bsnm{{Grummer-Strawn}},~\bfnm{Laurence~M.}\binits{L.~M.}},
\bauthor{\bsnm{Flegal},~\bfnm{Katherine~M.}\binits{K.~M.}},
\bauthor{\bsnm{Mei},~\bfnm{Zuguo}\binits{Z.}},
\bauthor{\bsnm{Wei},~\bfnm{Rong}\binits{R.}},
\bauthor{\bsnm{Curtin},~\bfnm{Lester~R.}\binits{L.~R.}},
\bauthor{\bsnm{Roche},~\bfnm{Alex~F.}\binits{A.~F.}} \AND
\bauthor{\bsnm{Johnson},~\bfnm{Clifford~L.}\binits{C.~L.}}
(\byear{2002}).
\btitle{2000 CDC growth charts for the United States: Methods and development}.
\bjournal{Vital and Health Statistics, Series 11}
\bvolume{246}.
\end{barticle}
%
\bptok{imsref}%
\endbibitem

\bibitem[\protect\citeauthoryear{Lambert and
Zheng}{2011}]{lambert:zheng:2011}
%
\begin{barticle}[author]
\bauthor{\bsnm{Lambert},~\bfnm{Peter}\binits{P.}} \AND
\bauthor{\bsnm{Zheng},~\bfnm{Buhong}\binits{B.}}
(\byear{2011}).
\btitle{On the consistent measurement of attainment and shortfall inequality}.
\bjournal{J. Health Econ.}
\bvolume{30}
\bpages{214--219}.
\end{barticle}
%
\bptok{imsref}%
\endbibitem

\bibitem[\protect\citeauthoryear{Langel and Till{\'
e}}{2013}]{langel:tille:2013}
%
\begin{barticle}[mr]
\bauthor{\bsnm{Langel},~\bfnm{Matti}\binits{M.}} \AND
\bauthor{\bsnm{Till{\'e}},~\bfnm{Yves}\binits{Y.}}
(\byear{2013}).
\btitle{Variance estimation of the {G}ini index: Revisiting a result
several times published}.
\bjournal{J. Roy. Statist. Soc. Ser. A}
\bvolume{176}
\bpages{521--540}.
\bid{doi={10.1111/j.1467-985X.2012.01048.x}, issn={0964-1998}, mr={3045858}}
\end{barticle}
%
\bptok{imsref}%
\endbibitem

\bibitem[\protect\citeauthoryear{Lumley}{2004}]{lumley:2004}
%
\begin{barticle}[author]
\bauthor{\bsnm{Lumley},~\bfnm{T.}\binits{T.}}
(\byear{2004}).
\btitle{Analysis of complex survey samples}.
\bjournal{J. Stat. Softw.}
\bvolume{9}
\bpages{1--19}.
\end{barticle}
%
\bptok{imsref}%
\endbibitem

\bibitem[\protect\citeauthoryear{Lumley}{2011}]{Rsurvey:2011}
%
\begin{bmisc}[author]
\bauthor{\bsnm{Lumley},~\bfnm{T.}\binits{T.}}
(\byear{2011}).
\bhowpublished{``Survey'': Analysis of complex survey samples.
R package version 3.26}.
\end{bmisc}
%
\bptok{imsref}%
\endbibitem

\bibitem[\protect\citeauthoryear{Lynch et~al.}{2004}]{lynch:etal:2004}
%
\begin{barticle}[author]
\bauthor{\bsnm{Lynch},~\bfnm{John}\binits{J.}},
\bauthor{\bsnm{Smith},~\bfnm{George~Davey}\binits{G.~D.}},
\bauthor{\bsnm{Harper},~\bfnm{Sam}\binits{S.}},
\bauthor{\bsnm{Hillemeier},~\bfnm{Marianne}\binits{M.}},
\bauthor{\bsnm{Ross},~\bfnm{Nancy}\binits{N.}},
\bauthor{\bsnm{Kaplan},~\bfnm{George~A.}\binits{G.~A.}} \AND
\bauthor{\bsnm{Wolfson},~\bfnm{Michael}\binits{M.}}
(\byear{2004}).
\btitle{Is income inequality a determinant of population health? Part
1. A systematic review}.
\bjournal{Milbank Q.}
\bvolume{82}
\bpages{5--99}.
\end{barticle}
%
\bptok{imsref}%
\endbibitem

\bibitem[\protect\citeauthoryear{Maasoumi}{1986}]{maasoumi:1986}
%
\begin{barticle}[author]
\bauthor{\bsnm{Maasoumi},~\bfnm{Esfandiar}\binits{E.}}
(\byear{1986}).
\btitle{The measurement and decomposition of multi-dimensional inequality}.
\bjournal{Econometrica}
\bvolume{54}
\bpages{991--998}.
\end{barticle}
%
\bptok{imsref}%
\endbibitem

\bibitem[\protect\citeauthoryear{Mackenbach and
Kunst}{1997}]{mackenbach:kunst:1997}
%
\begin{barticle}[pbm]
\bauthor{\bsnm{Mackenbach},~\bfnm{J.~P.}\binits{J.~P.}} \AND
\bauthor{\bsnm{Kunst},~\bfnm{A.~E.}\binits{A.~E.}}
(\byear{1997}).
\btitle{Measuring the magnitude of socio-economic inequalities in
health: An overview of available measures illustrated with two examples
from Europe}.
\bjournal{Soc. Sci. Med.}
\bvolume{44}
\bpages{757--771}.
\bid{issn={0277-9536}, pii={S0277953696000731}, pmid={9080560}}
\end{barticle}
%
\bptok{imsref}%
\endbibitem

\bibitem[\protect\citeauthoryear{Makdissi and Yazbeck}{2012}]{makdissi:yazbeck:2012}
%
\begin{bmisc}[author]
\bauthor{\bsnm{Makdissi},~\bfnm{Paul}\binits{P.}} \AND
\bauthor{\bsnm{Yazbeck},~\bfnm{Myra}\binits{M.}}
(\byear{2012}).
\bhowpublished{Avoiding blindness to health status:
A new class of health achievement and inequality indices.
Working Paper No. 1207E, Univ. Ottawa, Dept. Economics.
Available at \surl{http://www.sciencessociales.uottawa.ca/sites/\\default/files/public/eco/fra/documents/1207E.pdf}}.
\end{bmisc}
%
\bptok{imsref}%
\endbibitem

\bibitem[\protect\citeauthoryear{Mechanic}{2002}]{mechanic:2002}
%
\begin{barticle}[author]
\bauthor{\bsnm{Mechanic},~\bfnm{David}\binits{D.}}
(\byear{2002}).
\btitle{Disadvantage, inequality, and social policy: Major initiatives
intended to improve population health may also increase health disparities}.
\bjournal{Health Aff.}
\bvolume{21}
\bpages{48--59}.
\end{barticle}
%
\bptok{imsref}%
\endbibitem

\bibitem[\protect\citeauthoryear{Nakao and Treas}{1990}]{nakao:treas:1990}
%
\begin{bbook}[author]
\bauthor{\bsnm{Nakao},~\bfnm{Keiko}\binits{K.}} \AND
\bauthor{\bsnm{Treas},~\bfnm{Judith}\binits{J.}}
(\byear{1990}).
\btitle{Computing 1989 Occupational Prestige Scores}.
\bseries{GSS Methodology Reports}
\bvolume{70}.
\bpublisher{NORC},
\blocation{Chicago, IL}.
\end{bbook}
%
\bptok{imsref}%
\endbibitem

\bibitem[\protect\citeauthoryear{{NCHS}}{2011}]{fr:2011}
%
\begin{bbook}[author]
\bauthor{\bsnm{{NCHS}}}
(\byear{2011}).
\btitle{Healthy People 2010 Final Review}.
\bpublisher{National Center for Health Statistics (NCHS)},
\blocation{Hyattsville, MD}.
\end{bbook}
%
\bptok{imsref}%
\endbibitem

\bibitem[\protect\citeauthoryear{{O*NET}}{2014}]{onet:2014}
%
\begin{bmisc}[author]
\borganization{O*NET}
(\byear{2014}).
\bhowpublished{Onetcenter.org.
National Center for O*NET Development,
Washington, DC.
Available at \surl{http://www.onetcenter.org}.}
\end{bmisc}
%
\bptok{imsref}%
\endbibitem

\bibitem[\protect\citeauthoryear{Ogden et~al.}{2010}]{ogden:etal:2010}
%
\begin{barticle}[pbm]
\bauthor{\bsnm{Ogden},~\bfnm{Cynthia~L.}\binits{C.~L.}},
\bauthor{\bsnm{Lamb},~\bfnm{Molly~M.}\binits{M.~M.}},
\bauthor{\bsnm{Carroll},~\bfnm{Margaret~D.}\binits{M.~D.}} \AND
\bauthor{\bsnm{Flegal},~\bfnm{Katherine~M.}\binits{K.~M.}}
(\byear{2010}).
\btitle{Obesity and socioeconomic status in children and adolescents:
United States, 2005--2008}.
\bjournal{NCHS Data Brief}
\bvolume{51}
\bpages{1--8}.
\bid{issn={1941-4927}, pmid={21211166}}
\bptnote{check pages}%
\end{barticle}
%
\bptok{imsref}%
\endbibitem

\bibitem[\protect\citeauthoryear{Ogden et~al.}{2012}]{ogden:etal:2012}
%
\begin{barticle}[pbm]
\bauthor{\bsnm{Ogden},~\bfnm{Cynthia~L.}\binits{C.~L.}},
\bauthor{\bsnm{Carroll},~\bfnm{Margaret~D.}\binits{M.~D.}},
\bauthor{\bsnm{Kit},~\bfnm{Brian~K.}\binits{B.~K.}} \AND
\bauthor{\bsnm{Flegal},~\bfnm{Katherine~M.}\binits{K.~M.}}
(\byear{2012}).
\btitle{Prevalence of obesity in the United States, 2009--2010}.
\bjournal{NCHS Data Brief}
\bvolume{82}
\bpages{1--8}.
\bid{issn={1941-4927}, pmid={22617494}}
\bptnote{check pages}%
\end{barticle}
%
\bptok{imsref}%
\endbibitem

\bibitem[\protect\citeauthoryear{Pamuk}{1985}]{pamuk:1985}
%
\begin{barticle}[author]
\bauthor{\bsnm{Pamuk},~\bfnm{Elsie~R.}\binits{E.~R.}}
(\byear{1985}).
\btitle{Social class inequality in mortality from 1921 to 1972 in
England and {Wales}}.
\bjournal{Popul. Stud.}
\bvolume{39}
\bpages{17--31}.
\end{barticle}
%
\bptok{imsref}%
\endbibitem

\bibitem[\protect\citeauthoryear{Pamuk}{1988}]{pamuk:1988}
%
\begin{barticle}[author]
\bauthor{\bsnm{Pamuk},~\bfnm{Elsie~R.}\binits{E.~R.}}
(\byear{1988}).
\btitle{Social class inequality in infant mortality in England and
Wales from 1921 to 1980}.
\bjournal{European Journal of Population---Revue Europ\'{e}enne de D\'
{e}mographie}
\bvolume{4}
\bpages{1--22}.
\end{barticle}
%
\bptok{imsref}%
\endbibitem

\bibitem[\protect\citeauthoryear{Paruolo, Saisana and
Saltelli}{2013}]{paruolo:etal:2013}
%
\begin{barticle}[mr]
\bauthor{\bsnm{Paruolo},~\bfnm{Paolo}\binits{P.}},
\bauthor{\bsnm{Saisana},~\bfnm{Michaela}\binits{M.}} \AND
\bauthor{\bsnm{Saltelli},~\bfnm{Andrea}\binits{A.}}
(\byear{2013}).
\btitle{Ratings and rankings: Voodoo or science?}
\bjournal{J. Roy. Statist. Soc. Ser. A}
\bvolume{176}
\bpages{609--634}.
\bid{doi={10.1111/j.1467-985X.2012.01059.x}, issn={0964-1998}, mr={3067416}}
\end{barticle}
%
\bptok{imsref}%
\endbibitem

\bibitem[\protect\citeauthoryear{Peter}{2001}]{peter:2001}
%
\begin{barticle}[pbm]
\bauthor{\bsnm{Peter},~\bfnm{F.}\binits{F.}}
(\byear{2001}).
\btitle{Health equity and social justice}.
\bjournal{J. Appl. Philos.}
\bvolume{18}
\bpages{159--170}.
\bid{issn={0264-3758}, pmid={11785544}}
\end{barticle}
%
\bptok{imsref}%
\endbibitem

\bibitem[\protect\citeauthoryear{Pratt}{1964}]{pratt:1964}
%
\begin{barticle}[author]
\bauthor{\bsnm{Pratt},~\bfnm{John~W.}\binits{J.~W.}}
(\byear{1964}).
\btitle{Risk aversion in the small and the large}.
\bjournal{Econometrica}
\bvolume{32}
\bpages{122--136}.
\end{barticle}
%
\bptok{imsref}%
\endbibitem

\bibitem[\protect\citeauthoryear{{R Development Core Team}}{2011}]{R:2011}
%
\begin{bmisc}[author]
\borganization{R Development Core Team}
(\byear{2011}).
\bhowpublished{\textit{R: A Language and Environment for Statistical Computing}.
R Foundation for Statistical Computing,
Vienna, Austria}.
\end{bmisc}
%
\bptok{imsref}%
\endbibitem

\bibitem[\protect\citeauthoryear{Rao and Wu}{1988}]{rao:wu:1988}
%
\begin{barticle}[mr]
\bauthor{\bsnm{Rao},~\bfnm{J.~N.~K.}\binits{J.~N.~K.}} \AND
\bauthor{\bsnm{Wu},~\bfnm{C.-F.~J.}\binits{C.-F.~J.}}
(\byear{1988}).
\btitle{Resampling inference with complex survey data}.
\bjournal{J. Amer. Statist. Assoc.}
\bvolume{83}
\bpages{231--241}.
\bid{issn={0162-1459}, mr={0941020}}
\end{barticle}
%
\bptok{imsref}%
\endbibitem

\bibitem[\protect\citeauthoryear{Rao, Wu and Yue}{1992}]{rao:etal:1992}
%
\begin{barticle}[author]
\bauthor{\bsnm{Rao},~\bfnm{J.~N.~K.}\binits{J.~N.~K.}},
\bauthor{\bsnm{Wu},~\bfnm{C.~F.~J.}\binits{C.~F.~J.}} \AND
\bauthor{\bsnm{Yue},~\bfnm{K.}\binits{K.}}
(\byear{1992}).
\btitle{Some recent work in resampling methods}.
\bjournal{Surv. Methodol.}
\bvolume{18}
\bpages{209--217}.
\end{barticle}
%
\bptok{imsref}%
\endbibitem

\bibitem[\protect\citeauthoryear{Rawls}{1999}]{rawls:1971}
%
\begin{bbook}[author]
\bauthor{\bsnm{Rawls},~\bfnm{John~B.}\binits{J.~B.}}
(\byear{1999}).
\btitle{A Theory of Justice},
\bedition{Revised}~ed.
\bpublisher{Harvard Univ. Press},
\blocation{Cambridge, MA}.
\end{bbook}
%
\bptok{imsref}%
\endbibitem

\bibitem[\protect\citeauthoryear{R{\'e}nyi}{1961}]{renyi:1960}
%
\begin{bincollection}[mr]
\bauthor{\bsnm{R{\'e}nyi},~\bfnm{Alfr{\'e}d}\binits{A.}}
(\byear{1961}).
\btitle{On measures of entropy and information}.
In \bbooktitle{Proc. 4th {B}erkeley {S}ympos. {M}ath. {S}tatist. and
{P}rob., {V}ol. {I}}
\bpages{547--561}.
\bpublisher{Univ. California Press},
\blocation{Berkeley, CA}.
\bid{mr={0132570}}
\bptnote{check year}%
\end{bincollection}
%
\bptok{imsref}%
\endbibitem

\bibitem[\protect\citeauthoryear{Rose}{1985}]{rose:1985}
%
\begin{barticle}[author]
\bauthor{\bsnm{Rose},~\bfnm{Geoffrey}\binits{G.}}
(\byear{1985}).
\btitle{Sick individuals and sick populations}.
\bjournal{Int. J. Epidemiol.}
\bvolume{14}
\bpages{32--38}.
\end{barticle}
%
\bptok{imsref}%
\endbibitem

\bibitem[\protect\citeauthoryear{Rossen and Talih}{2014}]{rossen:talih:2014}
%
\begin{barticle}[pbm]
\bauthor{\bsnm{Rossen},~\bfnm{Lauren~M.}\binits{L.~M.}} \AND
\bauthor{\bsnm{Talih},~\bfnm{Makram}\binits{M.}}
(\byear{2014}).
\btitle{Social determinants of disparities in weight among US children
and adolescents}.
\bjournal{Ann. Epidemiol.}
\bvolume{24}
\bpages{705--713}.
\bid{doi={10.1016/j.annepidem.2014.07.010}, issn={1873-2585},
pii={S1047-2797(14)00360-3}, pmid={25174287}}
\bptnote{check pages}%
\end{barticle}
%
\bptok{imsref}%
\endbibitem

\bibitem[\protect\citeauthoryear{{RTI}}{2012}]{SUDAAN:2012}
%
\begin{bmisc}[author]
\borganization{RTI}
(\byear{2012}).
\bhowpublished{SUDAAN: Software for the Statistical Analysis of
Correlated Data, Release 11.
Research Triangle Institute (RTI),
Research Triangle Park, NC}.
\end{bmisc}
%
\bptok{imsref}%
\endbibitem

\bibitem[\protect\citeauthoryear{Saraiya et~al.}{2013}]{saraiya:etal:2013}
%
\begin{barticle}[author]
\bauthor{\bsnm{Saraiya},~\bfnm{Mona}\binits{M.}},
\bauthor{\bsnm{King},~\bfnm{Jessica}\binits{J.}},
\bauthor{\bsnm{Thompson},~\bfnm{Trevor}\binits{T.}},
\bauthor{\bsnm{Watson},~\bfnm{Meg}\binits{M.}},
\bauthor{\bsnm{Ajani},~\bfnm{Umed}\binits{U.}},
\bauthor{\bsnm{Li},~\bfnm{Jun}\binits{J.}} \AND
\bauthor{\bsnm{Houston},~\bfnm{Keisha~A.}\binits{K.~A.}}
(\byear{2013}).
\btitle{Cervical cancer screening among women aged 18--30
years---United States, 2000--2010}.
\bjournal{MMWR Morb. Mortal. Wkly. Rep.}
\bvolume{61}
\bpages{1038--1042}.
\end{barticle}
%
\bptok{imsref}%
\endbibitem

\bibitem[\protect\citeauthoryear{{SAS Institute}}{2010}]{SAS:2010}
%
\begin{bmisc}[author]
\borganization{SAS Institute}
(\byear{2010}).
\bhowpublished{SAS Proprietary Software 9.3.
SAS Institute Inc.,
Cary, NC}.
\end{bmisc}
%
\bptok{imsref}%
\endbibitem

\bibitem[\protect\citeauthoryear{{SEER Program}}{2013}]{SEERStat:2013}
%
\begin{bmisc}[author]
\borganization{SEER Program}
(\byear{2013}).
\bhowpublished{SEER*Stat Database: Incidence---SEER 18 Regs
Research Data${} + {}$Hurricane Katrina Impacted Louisiana Cases,
Nov 2012 Sub (2000--2010) $<$Katrina/Rita Population
Adjustment$>$---Linked to County Attributes---Total U.S.,
1969--2011 Counties.
National Cancer Institute, DCCPS, Surveillance Research Program,
Surveillance Systems Branch,
Bethesda, MD.
Released April 2013, based on the November 2012 submission.
Available at \surl{http://www.seer.cancer.gov}}.
\end{bmisc}
%
\bptok{imsref}%
\endbibitem

\bibitem[\protect\citeauthoryear{{Surveillance Research
Program}}{2013}]{SEERSoft:2013}
%
\begin{bmisc}[author]
\borganization{Surveillance Research Program}
(\byear{2013}).
\bhowpublished{SEER*Stat Software 8.1.2.
National Cancer Institute,
Bethesda, MD}.
\end{bmisc}
%
\bptok{imsref}%
\endbibitem

\bibitem[\protect\citeauthoryear{Talih}{2013a}]{talih:2013}
%
\begin{barticle}[pbm]
\bauthor{\bsnm{Talih},~\bfnm{Makram}\binits{M.}}
(\byear{2013}a).
\btitle{Invited commentary: Can changes in the distributions of and
associations between education and income bias estimates of temporal
trends in health disparities?}
\bjournal{Am. J. Epidemiol.}
\bvolume{177}
\bpages{882--884}.
\bid{doi={10.1093/aje/kwt042}, issn={1476-6256}, pii={kwt042}, pmid={23568596}}
\end{barticle}
%
\bptok{imsref}%
\endbibitem

\bibitem[\protect\citeauthoryear{Talih}{2013b}]{talih:2012}
%
\begin{barticle}[mr]
\bauthor{\bsnm{Talih},~\bfnm{Makram}\binits{M.}}
(\byear{2013}b).
\btitle{A reference-invariant health disparity index based on R\'enyi
divergence}.
\bjournal{Ann. Appl. Stat.}
\bvolume{7}
\bpages{1217--1243}.
\bid{doi={10.1214/12-AOAS621}, issn={1932-6157}, mr={3113507}}
\end{barticle}
%
\bptok{imsref}%
\endbibitem

\bibitem[\protect\citeauthoryear{Troiano and
Flegal}{1998}]{troiano:flegal:1998}
%
\begin{barticle}[author]
\bauthor{\bsnm{Troiano},~\bfnm{R.~P.}\binits{R.~P.}} \AND
\bauthor{\bsnm{Flegal},~\bfnm{Katherine~M.}\binits{K.~M.}}
(\byear{1998}).
\btitle{Overweight children and adolescents: Description,
epidemiology, and demographics}.
\bjournal{Pediatrics}
\bvolume{101}
\bpages{497--504}.
\end{barticle}
%
\bptok{imsref}%
\endbibitem

\bibitem[\protect\citeauthoryear{Tsui}{1999}]{tsui:1999}
%
\begin{barticle}[mr]
\bauthor{\bsnm{Tsui},~\bfnm{Kai-yuen}\binits{K.-y.}}
(\byear{1999}).
\btitle{Multidimensional inequality and multidimensional generalized
entropy measures: An axiomatic derivation}.
\bjournal{Soc. Choice Welf.}
\bvolume{16}
\bpages{145--157}.
\bid{doi={10.1007/s003550050136}, issn={0176-1714}, mr={1656440}}
\end{barticle}
%
\bptok{imsref}%
\endbibitem

\bibitem[\protect\citeauthoryear{Wagstaff}{2002}]{wagstaff:2002}
%
\begin{barticle}[pbm]
\bauthor{\bsnm{Wagstaff},~\bfnm{Adam}\binits{A.}}
(\byear{2002}).
\btitle{Inequality aversion, health inequalities and health achievement}.
\bjournal{J. Health Econ.}
\bvolume{21}
\bpages{627--641}.
\bid{issn={0167-6296}, pii={S0167-6296(02)00006-1}, pmid={12146594}}
\end{barticle}
%
\bptok{imsref}%
\endbibitem

\bibitem[\protect\citeauthoryear{Wagstaff}{2011}]{wagstaff:2011}
%
\begin{barticle}[pbm]
\bauthor{\bsnm{Wagstaff},~\bfnm{Adam}\binits{A.}}
(\byear{2011}).
\btitle{The concentration index of a binary outcome revisited}.
\bjournal{Health Econ.}
\bvolume{20}
\bpages{1155--1160}.
\bid{doi={10.1002/hec.1752}, issn={1099-1050}, pmid={21674677}}
\bptnote{check related}%
\end{barticle}
%
\bptok{imsref}%
\endbibitem

\bibitem[\protect\citeauthoryear{Wagstaff, Paci and van
Doorslaer}{1991}]{wagstaff:etal:1991}
%
\begin{barticle}[pbm]
\bauthor{\bsnm{Wagstaff},~\bfnm{A.}\binits{A.}},
\bauthor{\bsnm{Paci},~\bfnm{P.}\binits{P.}} \AND
\bauthor{\bparticle{van} \bsnm{Doorslaer},~\bfnm{E.}\binits{E.}}
(\byear{1991}).
\btitle{On the measurement of inequalities in health}.
\bjournal{Soc. Sci. Med.}
\bvolume{33}
\bpages{545--557}.
\bid{issn={0277-9536}, pmid={1962226}}
\end{barticle}
%
\bptok{imsref}%
\endbibitem

\bibitem[\protect\citeauthoryear{{WHO-CSDH}}{2008}]{who:2008}
%
\begin{bbook}[author]
\bauthor{\bsnm{{WHO-CSDH}}}
(\byear{2008}).
\btitle{Closing the Gap in a Generation: Health Equity Through Action
on the Social Determinants of Health. Final Report of the Commission on
Social Determinants of Health (CSDH)}.
\bpublisher{World Health Organization},
\blocation{Geneva}.
\end{bbook}
%
\bptok{imsref}%
\endbibitem

\bibitem[\protect\citeauthoryear{Wilson}{2009}]{wilson:2009}
%
\begin{barticle}[author]
\bauthor{\bsnm{Wilson},~\bfnm{James}\binits{J.}}
(\byear{2009}).
\btitle{Justice and the social determinants of health: An overview}.
\bjournal{Public Health Ethics}
\bvolume{2}
\bpages{210--213}.
\end{barticle}
%
\bptok{imsref}%
\endbibitem

\bibitem[\protect\citeauthoryear{Yin et~al.}{2010}]{yin:etal:2010}
%
\begin{barticle}[pbm]
\bauthor{\bsnm{Yin},~\bfnm{Daixin}\binits{D.}},
\bauthor{\bsnm{Morris},~\bfnm{Cyllene}\binits{C.}},
\bauthor{\bsnm{Allen},~\bfnm{Mark}\binits{M.}},
\bauthor{\bsnm{Cress},~\bfnm{Rosemary}\binits{R.}},
\bauthor{\bsnm{Bates},~\bfnm{Janet}\binits{J.}} \AND
\bauthor{\bsnm{Liu},~\bfnm{Lihua}\binits{L.}}
(\byear{2010}).
\btitle{Does socioeconomic disparity in cancer incidence vary across
racial/ethnic groups?}
\bjournal{Cancer Causes Control}
\bvolume{21}
\bpages{1721--1730}.
\bid{doi={10.1007/s10552-010-9601-y}, issn={1573-7225},
pmcid={2941051}, pmid={20567897}}
\end{barticle}
%
\bptok{imsref}%
\endbibitem

\bibitem[\protect\citeauthoryear{Young et~al.}{2001}]{staging:2001}
%
\begin{bbook}[author]
\bauthor{\bsnm{{Young}},~\bfnm{~L.}\binits{L.} \bsuffix{Jr.}},
\bauthor{\bsnm{Roffers},~\bfnm{S.~D.}\binits{S.~D.}},
\bauthor{\bsnm{Ries},~\bfnm{L.~A.~G.}\binits{L.~A.~G.}},
\bauthor{\bsnm{Fritz},~\bfnm{A.~G.}\binits{A.~G.}} \AND
\bauthor{\bsnm{Hurlbut},~\bfnm{A.~A.}\binits{A.~A.}}, eds.
(\byear{2001}).
\btitle{SEER Summary Staging Manual---2000: Codes and Coding Instructions}.
\bpublisher{National Cancer Institute},
\blocation{Bethesda, MD}.
\end{bbook}
%
\bptok{imsref}%
\endbibitem

\end{thebibliography}
%
%




\printaddresses
\end{document}